\documentclass[sigconf]{acmart}
\renewcommand\footnotetextcopyrightpermission[1]{} 
\usepackage{booktabs} 
\usepackage{threeparttable}

\usepackage{amsmath}
\usepackage{url}
\usepackage{graphicx}
\usepackage{multirow}
\usepackage{listings}
\usepackage{algorithm}
\usepackage{algorithmic}
\usepackage{subfigure}

\begin{document}
\title{Anomaly Analysis for Co-located Datacenter Workloads in the Alibaba Cluster}

\author{Rui Ren$^{*,+}$,Jinheng Li$^{\dagger}$,Lei Wang$^{*}$, Jianfeng Zhan$^{*}$, Zheng Cao$^{\star}$}
\affiliation{%
  \institution{$^{*}$State Key Laboratory of Computer Architecture, Institute of Computing Technology, CAS}
   \institution{$^{+}$University of Chinese Academy of Sciences}
   \institution{$^{\dagger}$ Sichuan University}
    \institution{$^{\star}$ Alibaba Co.,Ltd.}
}
\email{{renrui, lijinheng, wanglei_2011, zhanjianfeng}@ict.ac.cn, zhengzhi.cz@alibaba-inc.com}

\begin{abstract}
In warehouse-scale cloud datacenters, co-locating online services and offline batch jobs  is an efficient approach to improving
datacenter utilization.
To better facilitate the understanding of interactions among the co-located workloads and their real-world operational demands, Alibaba recently
released a cluster usage and co-located workload dataset, which is the first publicly  dataset with precise information about the category of each job.
In this paper, we perform a deep analysis on the released Alibaba workload dataset, from the perspective of anomaly analysis and diagnosis.
Through data preprocessing, node similarity analysis based on Dynamic Time Warping (DTW), co-located workloads characteristics analysis and anomaly analysis based on  iForest,
we reveals several insights including: (1) The performance discrepancy of machines in Alibaba's production cluster is relatively large, for
the distribution and resource utilization of co-located workloads is not balanced. For instance, the resource utilization (especially memory utilization) of  batch jobs  is fluctuating and  not as stable as that of online containers,
and the reason is that online containers are long-running jobs with more memory-demanding and most  batch jobs are short jobs,
(2) Based on the distribution of co-located workload instance numbers, the machines  can be classified into 8 workload distribution categories\footnote{\emph{workload distribution category} refers to the classification categories based on the number of batch tasks and online containers on machines.}.
And most patterns of machine resource utilization curves are similar in the same workload distribution category.
(3) In addition to the system failures, unreasonable scheduling and workload imbalance are the main causes of anomalies in Alibaba's cluster.
\end{abstract}

%
%


\maketitle

\section{Introduction}\label{intro}

With the popularity of internet services, cloud datacenter has become the infrastructure, which contains thousands of machines.
However, 
there is an IRU-QoS curse dilemma that  improving resource utilization (IRU) and guaranteeing QoS at the same time in cloud~\cite{Alibaba-Elasticity}~\cite{d2p}.
On one hand, in order to guarantee the service quality of internet services, the datacenter management systems usually reserve resources and it will
reduce the resource utilization.
For example, Geithner and McKinsey reported that the global server utilization seems to be very low, which is only 6\% to 12\%~\cite{Alibaba_cluster_trace}.
 Google reported that the CPU utilization of 20,000 servers averaged about 30\% during January to March, 2013, in a typical datacenter for online services~\cite{Datacenter_computer}.
 On the other hand, co-locating online services and offline batch jobs for resource sharing is an efficient approach to improving
datacenter utilization, even though it also raises unpredictable performance variability~\cite{d2p}.
 For instance, Alibaba tried to deploy batch jobs and latency-critical online services on same machines.  
They use Sigma~\cite{Alibaba_max} to schedule online service containers for the production jobs, and Fuxi~\cite{Fuxi} scheduler to manage the batch workloads. 
To better facilitate the understanding of interactions among the co-located workloads and their real-world operational demands, Alibaba
first released a co-located trace dataset (https://github.com/alibaba/clusterdata) in  Aug 2017.

For Alibaba's production cluster traces, recent studies~\cite{Alibaba_cluster_trace} \cite{Alibaba_Colocated_trace} \cite{Alibaba-Elasticity} have analyzed the characteristics from the perspective of imbalance phenomenon, co-located workloads (how the co-located
workloads interact and impact each other), the elasticity and plasticity of semi-containerized cloud.
However, these works do not further analyze the abnormal node in the cluster.
And discovering the cluster anomalies quickly is very important, for it  helps to locate bottlenecks, troubleshoot problems, avoid failures and improve utilization.

In this paper, we perform a deep analysis on the released Alibaba trace dataset~\cite{Alibaba_trace}, from the perspective of anomaly analysis and diagnosis. we first performed raw data preprocessing, including data supplementing, filtering, correlation and aggregation,
and generating the container-level, batch-level and server-level resource usage data finally.
Then based on these summary data, we conducted in-depth analysis from the aspects of node similarity, workload characteristics and distribution, and anomalies. 
From the above analysis, our key findings are summarized as follows:

\textbf{Performance discrepancy of machines in the Alibaba's co-located workloads cluster is relatively large}.
Obviously, The purpose of workloads co-locating  is making the resources which online services can not fully used dynamically to be fully used by batch jobs.
Unavoidably, deploying multiple applications to share resources on the same node will cause contentions and performance tilt.
In the Alibaba's cluster, the distribution of co-located workloads is not balanced. 
Since the online containers are long-running jobs with more memory-demanding and most  batch jobs are short jobs, the resource utilization (especially memory utilization) of  batch jobs  is fluctuating and  not as stable as that of online containers.
So the performance fluctuation between different nodes may be high.

\textbf{Generally, the patterns of machine resource utilization curves are similar in the same workload distribution category}.  Based on the co-located workload distributions, the machines of Alibaba's cluster can be classified into 8 categories.
Especially, most of CPU usage and memory usage on machines that belonging to the same category have similar patterns, while disk usage of different nodes may vary greatly.

\textbf{Unreasonable scheduling and workload imbalance are the main causes of anomalies in Alibaba's cluster}.
Undoubtedly, system errors or failures  will cause the abnormal nodes or  unavailable nodes.
In addition, the scheduling strategies of the cluster management systems are unreasonable, may result in uneven workload distribution.
And the resource contentions or interferences from co-located workloads, may also cause  abnormal resource utilization.
Since the Alibaba cluster is a co-located workloads datacenter, the abnormal phenomenons  caused by  imbalance workload and utilization are more
common.

\section{Background and Methodology  }

\subsection{Trace Overview}

In Aug 2017, Alibaba released a publicly accessible trace data referred as ``Alibaba Cluster Trace''.
The trace data contains cluster information of a production cluster in 12 hours period, and contains about 1.3k machines that run both online services and offline batch jobs. And the  dataset includes six files: \emph{server\_event.csv}, \emph{server\_usage.csv}, \emph{batch\_instance.csv},  \emph{batch\_task.csv}, \emph{container\_event.csv} and \emph{container\_usage.csv},
which can be classified into two categories: 1) resource data, and 2) workload data.

\subsubsection{Resource Data}

In the Alibaba cluster, Alibaba CMS (cluster management system) provides a practice of semi-containerized co-location:  online services
run in containers and  batch jobs directly run on physical servers~\cite{Alibaba-Elasticity}.
So the dataset includes the resource utilization data on physical machines and containers.

\paragraph{Physical Machine Resource Usage}

The resource utilization information of physical machine includes two files:  \emph{server\_event.csv} and \emph{server\_usage.csv}.

The file \emph{server\_event.csv} reflects the normalized physical capacity of each machine and event type~\cite{schema}.
It gives three dimensions of physical capacity: CPU cores, Memory and Disk, and each dimension is normalized independently.
The three  event types are \emph{add}, \emph{softerror} and \emph{harderror}. In total, there are 1313 64-cores machines in the Alibaba Cluster Trace, whose machine id is from 1 to 1313.

The file \emph{server\_usage.csv} reflects the total resource utilization of all workloads (batch tasks, online container instances and the workloads of operating systems) in physical machines. It records the CPU usage, memory  usage, disk usage and the average linux cpu load of 1/5/15 minute during
a period from 39600s  to 82500s\footnote{The  timestamp of each record in the trace,  which are in seconds and relative to the start of trace period. Additionally, a time of 0 represents the event occur before the trace period~\cite{Alibaba_trace}.}, and the most of the record time interval is 300s.
However, the data in file \emph{server\_usage.csv} is partially  missing. For example, it only records the resource utilization data of 1310 machines.

\paragraph{Container Resource Usage}

The resource utilization information of online container also includes two files:  \emph{container\_event.csv} and \emph{container\_usage.csv}.


The file \emph{container\_event.csv} reflects the created online containers, and their requested resources, which including the assigned CPU cores, memory and disk space.
Since the instance is the smallest scheduling unit and running in a lightweight virtual machine of Linux container (LXC), each instance can be seen as a container. In addition, it could also be regarded as a online service job.
From the file  \emph{container\_event.csv}  we see that, there is just one event type in container instance, which is \emph{Create}.
That is, the online container instance will always exist if it is not killed after being created, which can be consider as a long-running job workload.

The file \emph{container\_usage.csv} gives the actual resource utilization information for online container instances, such as, cpu usage, memory usage, disk usage, average cpu load, cache misses.
Most container resource utilization data is also collected from  39600s to  82500s, and the measurement interval of resource usage data  is  about 300s.

\subsubsection{Workload Data}

There are two files \emph{batch\_task.csv} and \emph{batch\_instance.csv} to describe the batch workloads.
In general, a batch job contains multiple tasks, and different tasks execute different computing logics according to the data dependencies.
In addition, a task may be executed through multiple instances, which execute exactly the same binary with the same resource request, but with different input data~\cite{Alibaba_cluster_trace}.
So the file \emph{batch\_task.csv} describes  the task execution information of batch jobs, such as, the task status, the cpu and memory resources that tasks require. 

The file \emph{batch\_instance.csv} also gives the batch instance information, which is the smallest scheduling unit of batch workload. 
In addition,a batch instance may fail due to machine failures or network problems. Each record in this file records one try run. The start and end timestamp can be 0 for some instances. For example, all timestamp is zero when
 the instance is in ready and waiting status; start time is non-zero but end time is zero, when the instance is in failed status.

\subsection{Our Methodology}

Based on the Alibaba Cluster Trace,  researchers can study the workload characteristics, analyze the cluster status,
design new algorithms to assign workloads, and help to optimize the scheduling strategies between online services and batch jobs for improving throughput while maintain acceptable service quality.
For large-scale clusters, anomaly discovery and diagnosis is also very important, which  helps to locate bottlenecks, troubleshoot problems, avoid failures and improve utilization. 
In addition, it is a common and effective method to perform anomaly analysis from trace dataset.

\begin{figure}[t]
   \begin{center}
   \includegraphics[scale=0.36]{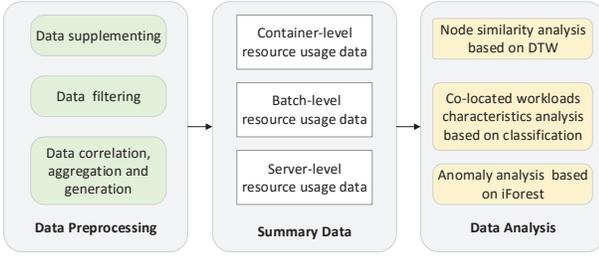}
   \end{center}
    \caption{The analysis methodology.}
    \label{method}
   \end{figure}

In this paper, we perform a deep analysis on the released Alibaba trace dataset, from a distinctive perspective of anomaly analysis and diagnosis.
The analysis method is shown in Figure~\ref{method}.
We first performed raw data preprocessing, including data supplementing, filtering, correlation and aggregation,
and generating the container-level, batch-level and server-level resource usage data finally.
Then based on these summary data, we conducted in-depth analysis from the aspects of node similarity, workload characteristics and distribution, and anomalies.

\subsection{Terminology}\label{terminology}

To analyze the machine conditions and discover anomalies in the Alibaba cluster , we correlate the multiple files and define the following symbols:
\begin{itemize}
\item  $m$: The machine id of the cluster, which is range from 1 to 1313.

\item  $ci$: The container instance, whose start time and end time are $t(ci)_{start}$ and $t(ci)_{end}$ , respectively.

\item  $bi$: The batch task instance, whose start time and end time are $t(bi)_{start}$ and $t(bi)_{end}$, respectively.


\item $t_{x}$: The $x^{th}$ recording timestamp.

\item $I_{x}$: The $x^{th}$  time interval, here, $I_{x}=[t_{x},t_{x+1}]$.

\item $CpuNum_{m}$: The number of CPU cores that machine $m$ has.

\item $CpuNum(ci)^{req}$: The number of  CPU cores that requested by the container instance $ci$.

\item $CpuNum(bi)^{req}$: The number of  CPU cores that requested by the batch task instance $bi$.

\item $Mem(ci)^{reg}$: The memory that requested by the container instance $ci$.

\item $CpuPercent(ci)^{used}$: The used percent of container instance $ci$'s requested cpus.

\item $CpuNum(ci)^{used}_{m}$: The number of CPU core that container instance $ci$  used on machine $m$.

\item $CpuNum(bi)^{used}_{m}$: The number of CPU core that  batch task instance $bi$ used on machine $m$.
\end{itemize}

In the subsequent analysis, we will summarize the trace data during every time interval,
so we also define the following symbols to describe the resource utilization  during the time interval $I_{x}$:

\begin{itemize}
\item $Set(ci)_{m,I_{x}}$: The online container instance data sets that on machine $m$, and their life cycles have intersections with the time interval $I_{X}$.
    
\item   $Set(bi)_{m,I_{x}}$: The  batch task instance data sets that on machine $m$, and their life cycles have intersections with the time interval $I_{X}$.

\item $Num(ci)_{m,I_{x}}$: The number of container instance $ci$  that running on machine $m$.

\item $Num(bi)_{m,I_{x}}$: The number of  batch task instance $bi$ that running on machine $m$.

\item $CpuNum(bi)^{used}_{m,I_{x}}$: The estimate number of CPU cores that container instance $ci$ or batch task $bi$. 

\item $CpuPercent(ci)^{used}_{m,I_{x}}$: The proportion of used CPU resources in the requested CPU resources of container $ci$. 

\item $MemPercent(ci)^{used}_{m,I_{x}}$: The proportion of used memory in the requested memory resources of container $ci$.

\item $CpuUsage(ci)_{m,I_{x}}$:  The actual CPU usage of container instance $ci$  that running on machine $m$.

\item $CpuUsage(bi)_{m,I_{x}}$:  The actual CPU usage of  batch task instance $bi$ that running on machine $m$.

\item  $MemUsage(ci)_{m,I_{x}}$:  The actual memory usage of container instance $ci$  that running on machine $m$.

\item  $MemUsage(bi)_{m,I_{x}}$:  The actual memory usage of  batch task instance $bi$ that running on machine $m$.

\item $RT(bi)_{m,I_{x}}$: The real occupation runtime of batch task instance $bi$ that running on machine $m$.

\item $Total\_CpuNum(ci)_{m,I_{x}}$: The total number of CPU core that container instance $ci$  used on machine $m$.
    
\item $Total\_CpuNum(bi)_{m,I_{x}}$: The total number of CPU core that batch task $bi$ used on machine $m$.

\item $Runtime(bi)_{m}$: The runtime of batch task instance $bi$ that running on machine $m$.

\item $Total\_CpuUsage(ci)_{m,I_{x}}$:  The total CPU usage of all container instances  instances that running on machine $m$.
    
\item $Total\_CpuUsage(bi)_{m,I_{x}}$:  The total CPU usage of all  batch task instances that running on machine $m$.

\item $Total\_MemUsage(ci)_{m,I_{x}}$:   The total memory usage of all container instances  that running on machine $m$.
    
\item $Total\_MemUsage(bi)_{m,I_{x}}$:   The total memory usage of all batch task  instances that running on machine $m$.

\item  $CpuUsage_{m,I_{x}}$: The average CPU usage of machine $m$.

\item  $MemUsage_{m,I_{x}}$: The average memory usage of machine $m$.

\item  $DiskUsage_{m,I_{x}}$: The average disk usage of machine $m$.

\end{itemize}

\section{Data Preprocessing}

\subsection{Data Supplementing}

We find that some files have missing data.
Such as, there is no resource data of three machines (machine id is \emph{149}, \emph{602} and \emph{930}) in the file \emph{server\_usage.csv}.
And it samples the resource usage of each machine every 300s from 39600s to 82500s.
That is, 144 resource utilization data is sampled at each machine.
In fact, we find that on 335 machines, the number of recorded resource utilization data is less than 144,
which means the resource data of some machines are missing, too.

So we do the data supplementing.
For the missing machine \emph{149}, \emph{602} and \emph{930}, all resource data is completed with 0.	
Afterwards, the linear interpolation method is used to replenish the data, which is a method of constructing new data points within the range of a discrete set of known data points~\cite{Interpolation}.
For example, supposing $a(i)$ is the $i^{th}$ missing data, and the number of missing value between existing value $b$ and $b^{'}$ is $Num_{miss}$.
And the detailed linear interpolation method is described in Algorithm~\ref{interpolation}.

\begin{algorithm}[tp]
\footnotesize
\renewcommand{\algorithmicrequire}{ \textbf{Input:}} 
\renewcommand{\algorithmicensure}{ \textbf{Output:}} 
\caption{\footnotesize Linear interpolation method.}
\label{interpolation}
\begin{algorithmic}[1]
\STATE Find $b$ and $b^{'}$
\STATE Calculate the $Num_{miss}$ between  $b$ and $b^{'}$
\STATE Calculate the rake ratio $rake\_ratio$ between $b$ and $b^{'}$：
\STATE $rake\_ratio=\frac{(b^{'}-b)}{Num_{miss}-1}$
\FOR {each miss value $a(i)$ }
  \STATE  $a(i)=b+rake\_ratio*i$
  \STATE Insert $a(i)$ into the raw data
 \ENDFOR
\end{algorithmic}
\end{algorithm}



\subsection{Data Filtering}

Some files also have the aberrant data, which needs to be deleted.
 For example, the record number of \emph{container\_event.csv} is 11102, while the number of online container instances  that we calculated is just 11089.
Through our in-depth analysis, we find that some online container instances  are duplicated and have two memory allocation values, which are shown in Table~\ref{ab_mem}.
We can see that, at the same time, there are multiple containers on the same node. If the requested memory of one container is greater than 0.9,
all the requested memory of containers may be exceed the machine memory, which is obviously unreasonable.
So we remove these anomalous records that requested memory is greater than 0.9.


\begin{table}[htpb]
\renewcommand{\arraystretch}{1.3}
\footnotesize
\centering
\caption{The data with abnormal requested memory in \emph{container\_event.csv}.}
\begin{tabular}{|p{1cm}|p{3cm}|p{1.3cm}|p{1.3cm}|}
  \hline
  $ci$ & $Mem(ci)^{req}$  & $m$  &  $Num(ci)_{m}$ \\ \hline
1681 &  0.0424093 / 0.999963 & \emph{56} & 10 \\
2160 &  0.0424093 / 1 & \emph{1038} &  9 \\
2878 & 0.0424093 / 1  & \emph{1112} & 8 \\
3384 & 0.0848187 / 0.999963 & \emph{102} & 8  \\
4467 &  0.0424093 /1    & \emph{331}  &  10 \\
5470 &  0.0424093 /0.999963  & \emph{866} & 9  \\
6330 &   0.0424093 /  0.999963 & \emph{95} & 8 \\
6549 &  0.0848187 / 1.00001 &  \emph{1134} & 9 \\
6639 &   0.0424093  / 0.999963 & \emph{19} & 9 \\
7663 &  0.0424093 / 1 & 323 &  \emph{10} \\
7915 &  0.0424093 / 0.999963 & \emph{69} &  12 \\
8476 &  0.0424093 /1 & \emph{323} & 10 \\
10772 &  0.0424093 /  0.999963 & \emph{85} & 12 \\
  \hline
\end{tabular}
\label{ab_mem}
\end{table}

\subsection{Data Correlation, Aggregation and Generation}

In order to compare the resource utilization of  online container services, batch job workloads and servers, we aggregate all the container-level, batch-level and server-level resource usage statistics by the machine Id and recording interval, respectively.


\subsubsection{Generating container-level resource usage data}


Because the file \emph{container\_usage.csv} samples the resource usage of each container every 300s.
So at every time interval, we aggregate all the container-level resource usage statistics by machine Id based on $ container \rightarrow machine\_Id$ mapping recorded in the \emph{container\_event.csv}~\cite{Alibaba_cluster_trace}. We generate the container instance data sets $Set(ci)_{m,I_{x}}$.
And then, the CPU usage and memory usage that occupied by all containers during every interval is defined as $Total\_CpuUsage(ci)_{m,I_{x}}$ and $Total\_MemUsage(ci)_{m,I_{x}}$, which can be calculated by Algorithm~\ref{cpu_container}.

\begin{algorithm}[tp]
\footnotesize
\renewcommand{\algorithmicrequire}{ \textbf{Input:}} 
\renewcommand{\algorithmicensure}{ \textbf{Output:}} 
\caption{\footnotesize 	Calculating the CPU usage and memory usage of all containers during every interval.}
\label{cpu_container}
\begin{algorithmic}[1]
\REQUIRE ~~\\
 $Set(ci)_{m,I_{x}}$
\ENSURE ~~\\
$Total\_CpuUsage(ci)_{m,I_{x}}$,$Total\_MemUsage(ci)_{m,I_{x}}$
\STATE Select the online container instances set $Set(ci)_{m,I_{x}}$ on machine $m$ within $I_{x}$
\STATE Count the $Num(ci)_{m,I_{x}}$
\FOR {each online container instance in $Set(ci)_{m,I_{x}}$}
  \STATE Calculate $CpuUsage(ci)_{m,I_{x}}$:
  \STATE $CpuUsage(ci)_{m,I_{x}}$=$\frac{CpuPercent(ci)^{used}_{m,I_{x}} * CpuNum(ci)^{req}}{CpuNum_{m}}$
  \STATE Calculate  $MemUsage(ci)_{m,I_{x}}$:
  \STATE $MemUsage(ci)_{m,I_{x}}$=$MemPercent(ci)^{used}_{m,I_{x}}$*$Mem(ci)^{req}$
 \ENDFOR
 \STATE $Total\_CpuUsage(ci)_{m,I_{x}}= \sum CpuUsage(ci)_{m,I_{x}}$
  \STATE $Total\_MemUsage(ci)_{m,I_{x}}= \sum MemUsage(ci)_{m,I_{x}}$
\end{algorithmic}
\end{algorithm}

\subsubsection{Generating batch-level resource usage data}

Cheng et al.~\cite{Alibaba_Colocated_trace} have calculated the batch job workload resource usage  by subtracting the usage of containers
from the overall usage of the cluster. However, we think their calculation method is not accurate enough, for there are resources that occupied by the OS operations on  machines, except for the resources used by containers and  batch tasks.
So we generate the batch-level resource usage data based on  actual occupation time of batch task instances.

\begin{figure}
   \begin{center}
   \includegraphics[scale=0.8]{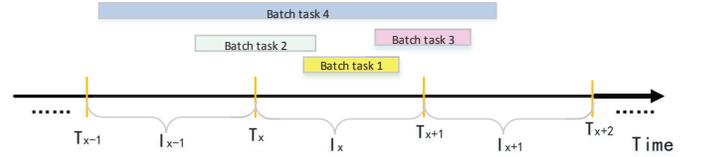}
   \end{center}
    \caption{Four situations for the positions of batch tasks' start and end time.}
    \label{batch_task_runtime}
   \end{figure}

The file \emph{batch\_instance.csv} records the start time, end time and location (machine) of all batch task instances.
For each time interval, according to the positions of  batch tasks' start time and end time, there are four situations that shown in Figure~\ref{batch_task_runtime}.
 So we can  calculate the actual occupation time of batch task instances during every time interval, according to formula~(\ref{batch_runtime}).

\begin{center}
\begin{equation}
\scriptsize
\label{batch_runtime}
\left\{
 \begin{array}{lr}
  RT(bi)_{m,I_{x}}=t(bi)_{end} - t(bi)_{start}  \quad  (t(bi)_{start} \geq t_{x} ~ \& ~ t(bi)_{end} \leq t_{x+1})  \\
  RT(bi)_{m,I_{x}} =t(bi)_{end} - t_{x}  \quad  (t(bi)_{start} \leq t_{x} ~ \&  ~ t(bi)_{end} \leq t_{x+1}) \\
  RT(bi)_{m,I_{x}}=t_{x+1} - t(bi)_{start} \quad   (t(bi)_{start} \geq t_{x} ~ \&  ~ t(bi)_{end} \geq t_{x+1}) \\
  RT(bi)_{m,I_{x}}=t_{x+1} - t_{x} \quad  (t(bi)_{start} \leq t_{x} ~ \&  ~ t(bi)_{end} \geq t_{x+1}) &
 \end{array}
\right.
\end{equation}
\end{center}

So we derive the cpu usage that occupied by all batch tasks based on the task execution time at every time interval.
And the CPU usage and memory usage that occupied by all batch tasks during every interval is $Total\_CpuUsage(bi)_{m,I_{x}}$ and $Total\_MemUsage(bi)_{m,I_{x}}$, which can be calculated by Algorithm~\ref{usage-bacth}.


\subsubsection{Generating server-level resource usage data}

Similarly, based on the file \emph{server\_usage.csv}, we calculate the average resource utilization for each time interval and each machine, which includes $CpuUsage_{m,I_{x}}$, $MemUsage_{m,I_{x}}$ and $DiskUsage_{m,I_{x}}$.

After generating the above data, a series of analysis can be performed on the basis of server-level, container-level and batch-level resource usage data, such as, node similarity analysis, co-located workloads characteristics analysis, and anomaly analysis, and so on.



\begin{algorithm}[tp]
\linespread{1.5}
\footnotesize
\renewcommand{\algorithmicrequire}{ \textbf{Input:}} 
\renewcommand{\algorithmicensure}{ \textbf{Output:}} 
\caption{\footnotesize 	Calculating the CPU usage and memory usage of all batch tasks during every interval.}
\label{usage-bacth}
\begin{algorithmic}[1]
\REQUIRE ~~\\
$set_{m,x}(bi)$
\ENSURE ~~\\
$Total\_CpuUsage(bi)_{m,I_{x}}$, $Total\_MemUsage(bi)_{m,I_{x}}$
\STATE Select the batch instances set $set(bi)_{m,x}$ within $I_{i}$
\STATE Calculate the  $Num(bi)_{m,I_{x}}$
\FOR {each batch instance in $set(bi)_{m,I_{x}}$}
  \IF {$t(bi)_{start} \geq  t_{x}$  and $t(bi)_{end} \leq t_{x+1}$}
  \STATE $RT(bi)_{m,I_{x}}=t(bi)_{end} - t(bi)_{start}$
  \STATE $CpuNum(bi)^{used}_{m,I_{x}}=CpuNum(bi)^{used}_{m}$
  \STATE $Mem(bi)^{used}_{m,I_{x}}=Mem(bi)^{used}_{m}$
  \ELSIF {$t(bi)_{start} \leq t_{x}$ and $t(bi)_{end} \leq t_{x+1}$}
  \STATE $RT(bi)_{m,I_{x}} =t(bi)_{end} - t_{x}$
  \STATE $CpuNum(bi)^{used}_{m,I_{x}}$=$\frac{RT(bi)_{m,I_{x}}}{RT(bi)_{m}}$*$CpuNum(bi)^{used}_{m}$
   \STATE $Mem(bi)^{used}_{m,I_{x}}$=$\frac{RT(bi)_{m,I_{x}}}{RT(bi)_{m}}$*$Mem(bi)^{used}_{m}$
    \ELSIF {$t(bi)_{start} \geq t_{x}$ and $t(bi)_{end} \geq t_{x+1}$}
  \STATE $RT(bi)_{m,I_{x}}=t_{x+1} - t(bi)_{start}$
  \STATE $CpuNum(bi)^{used}_{m,I_{x}}$=$\frac{RT(bi)_{m,I_{x}}}{RT(bi)_{m}}$*$CpuNum(bi)^{used}_{m}$
   \STATE $Mem(bi)^{used}_{m,I_{x}}$=$\frac{RT(bi)_{m,I_{x}}}{RT(bi)_{m}}$*$Mem(bi)^{used}_{m}$
    \ELSIF {$t(bi)_{start} \leq t_{x}$ and $t(bi)_{end} \geq t_{x+1}$}
  \STATE $RT(bi)_{m,I_{x}}=t_{x+1} - t_{x}$
  \STATE $CpuNum(bi)^{used}_{m,I_{x}}$=$\frac{RT(bi)_{m,I_{x}}}{RT(bi)_{m}}$*$CpuNum(bi)^{used}_{m}$
   \STATE $Mem(bi)^{used}_{m,I_{x}}$=$\frac{RT(bi)_{m,I_{x}}}{RT(bi)_{m}}$*$Mem(bi)^{used}_{m}$
  \ENDIF
 \ENDFOR
\STATE $Total\_CpuNum(bi)_{m,I_{x}}= \sum CpuNum(bi)^{used}_{m,I_{x}} $
 \STATE $Total\_CpuUsage(bi)_{m,I_{x}}= \frac{Total\_CpuNum(bi)_{m,I_{x}}}{CpuNum_{m}} $
  \STATE $Total\_MemUsage(bi)_{m,I_{x}}= \sum Mem(bi)^{used}_{m,I_{x}} $
\end{algorithmic}
\end{algorithm}

\section{Node Similarity Analysis}\label{similarity}

Node similarity analysis can be used to discover the  performance difference between nodes in the cluster, and help to understand the stability of cluster. In this section, we apply Dynamic Time Warping (DTW)~\cite{DTW} to measure the similarity between server-level resource utilization series.

\subsection{Node similarity Analysis based on DTW}


\subsubsection{Calculating DTW value between two time series}

Dynamic Time Warping (DTW) is a distance measure that  compares two time series after optimally aligning them.
Suppose there are two time series $Q$ and $S$, of length $n$ and $l$ respectively, where:

\begin{equation} \label{sequence1}
\footnotesize
Q=(q_{1}, q_{2},...,q_{i},...,q_{n})
\end{equation}
\begin{equation} \label{sequence2}
\footnotesize
S=(s_{1}, s_{2},...,s_{j},...,s_{l})
\end{equation}

Then we construct an n-by-l matrix, and  the ($i^{th}$,
$j^{th}$) element of the matrix contains the distance $d(q_{i},s_{j})$ between the two points $q_{i}$ and $c_{j}$ ( Typically, the Euclidean distance is used, so $d(q_{i},s_{j}) = (q_{i} - s_{j})^2 $).
Each matrix element $(i,j)$ corresponds to the alignment between the points $q_{i}$ and $s_{j}$. 
The warping path $W$ is a contiguous  set of matrix elements that defines a mapping between $Q$ and $S$. So The $k^{th}$ element of $W$ is defined as $w_{k} = (i,j)_{k}$, and:
 \begin{equation} \label{W}
\footnotesize
W=(w_{1}, w_{2},...,w_{k}, w_{K})  ~~~~  max(l,n) \leq K \le l+n+1
\end{equation}

In addition, we are interested  in the path which minimizes the warping cost:
 \begin{equation} \label{DTW}
\footnotesize
DTW(Q,S)= min \left\{ \frac{1}{K} \sqrt{\sum_{k=1}^K w_{k}}	\right\}
\end{equation}
The $K$ in the denominator is used to compensate for the fact that warping paths may have
different lengths. This path can be found very efficiently using dynamic programming, and
we  define  the cumulative distance $dtw(i,j)$, as
the distance $d(q_{i},s_{j})$ found in the current cell and the minimum of the cumulative distances of the adjacent elements~\cite{DTW_series}:
\begin{equation} \label{DTW}
\footnotesize
dtw(i,j)= d(q_{i},s_{j})+ min \left\{dtw(i-1,j-1),dtw(i-1,j),dtw(i,j-1)\right\}
\end{equation}

\begin{figure*}[htbp]
\begin{minipage}[t]{0.32\linewidth}
\centering
\includegraphics[height=3cm,width=5cm]{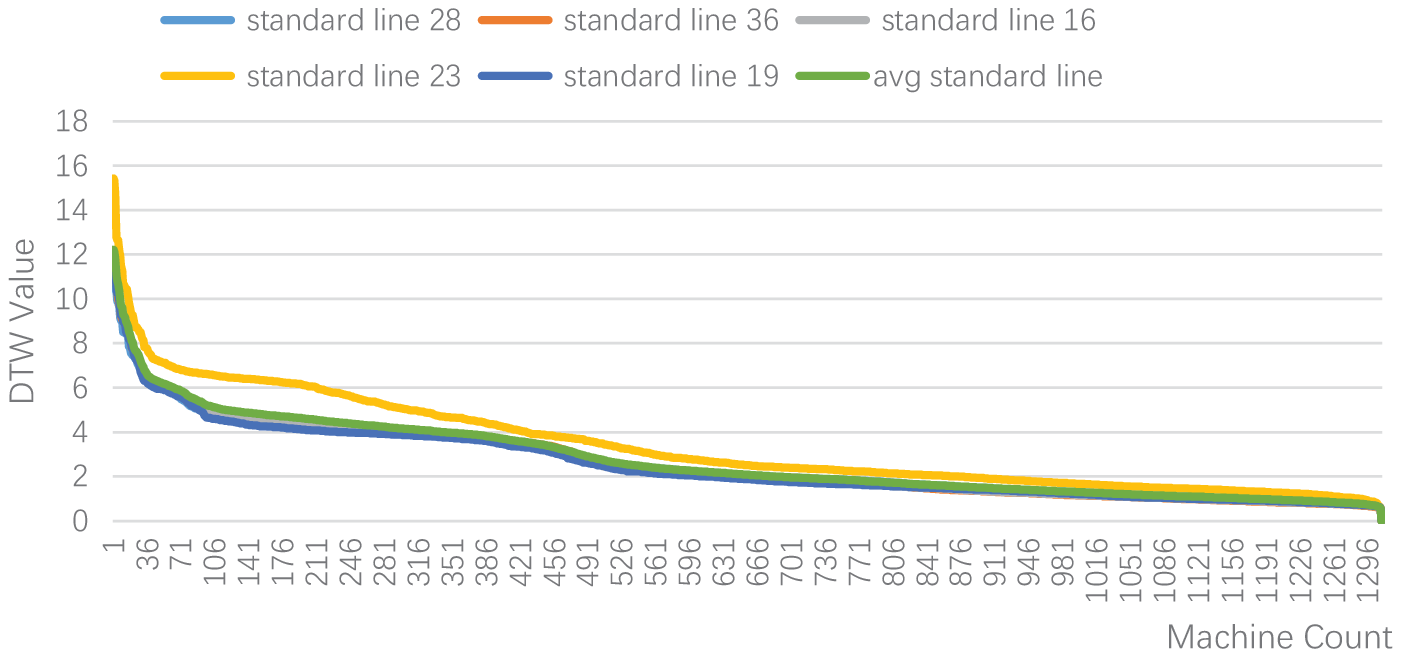}
\caption{Sorted DTW values.}
 \label{dtw-machine}
\end{minipage}%
\begin{minipage}[t]{0.33\linewidth}
\centering
\includegraphics[height=3cm,width=5.2cm]{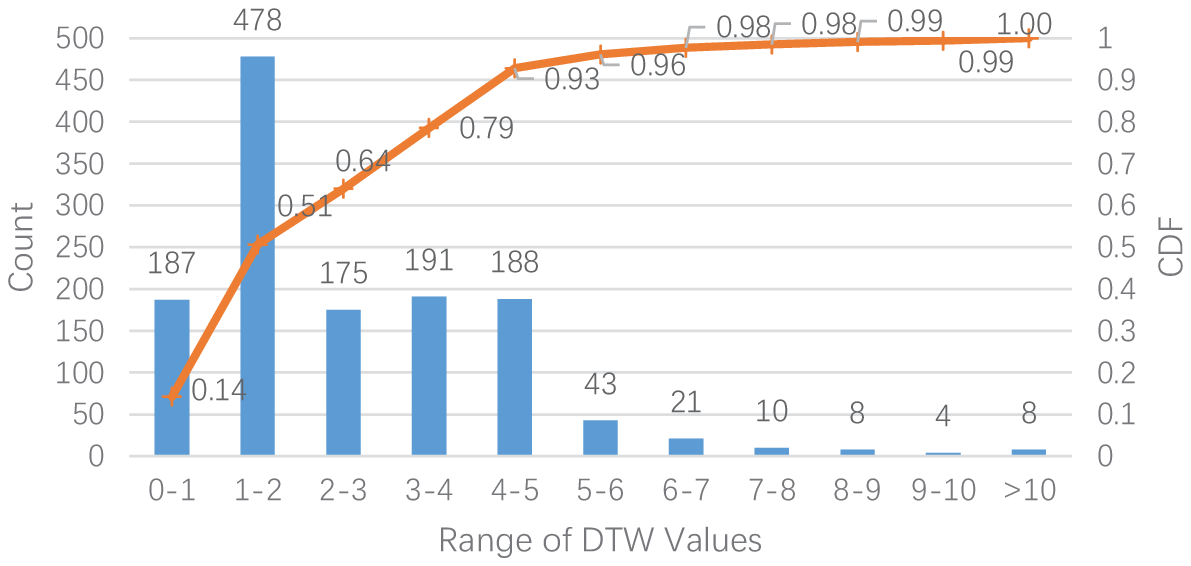}
\caption{DTW ranges.}
 \label{dtw-range}
\end{minipage}
\begin{minipage}[t]{0.33\linewidth}
\centering
\includegraphics[height=3cm,width=6cm]{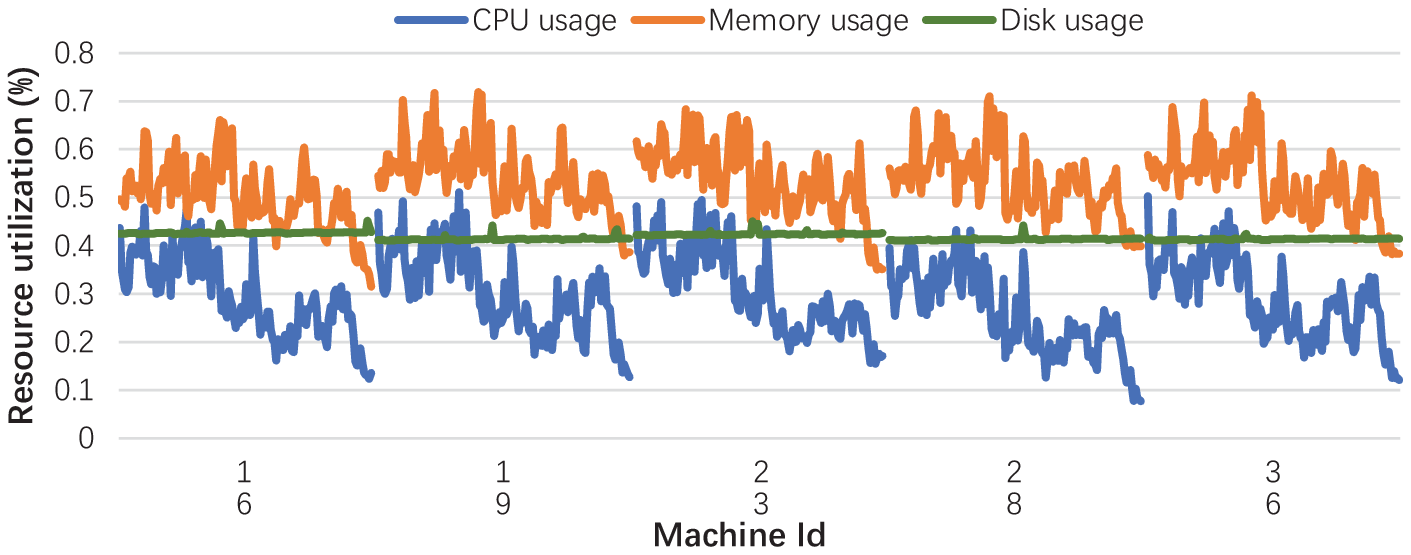}
\caption{Resource utilization of the selected standard curves.}
\label{standard-curve}
\end{minipage}%
\end{figure*}

\subsubsection{Calculating node similarity based DTW value}

During the tracing interval, the resource utilization on each machine can form a resource utilization curve.
Based on the \emph{server\_usage.csv} that has been supplemented, we combine these three curves of CPU usage, memory usage and disk usage into a resource utilization curve.
All the resource  utilization curves constitute a data set $Set(resource\_curve)$.
Then, we try to find a standard curve $standard\_curve$ by random sampling, and calculate the DWT values between  all other resource curves and the standard curve, which is taken as the \emph{node similarity}.

\begin{algorithm}[tp]
\footnotesize
\renewcommand{\algorithmicrequire}{ \textbf{Input:}} 
\renewcommand{\algorithmicensure}{ \textbf{Output:}} 
\caption{\footnotesize Calculating the DTW value of the cluster nodes.}
\label{DTW-detection}
\begin{algorithmic}[1]
\REQUIRE ~~\\
$Set(resource\_curve)$
\ENSURE ~~\\
DTW values
\STATE Randomly extract the $sample\_num$ rows in $Set(resource\_curve)$,
   and set the obtained rows as the sample set $Set(sample)$.
\FOR {each row in $Set(sample)$}
  \STATE  Calculate the DTW value in each pair
  \STATE  Put the calculated DTW value into the array $DTW(sample)$
 \ENDFOR
  \STATE Take the median value of  $DTW(sample)$ as the standard value 
  \STATE  Select a row as the standard curve $standard\_curve$ randomly.
  \FOR {each row in $Set(resource\_curve)$}
  \STATE Calculate the DTW value of each row between $standard\_curve$ 
   \ENDFOR
\end{algorithmic}
\end{algorithm}

\subsection{The results of node similarity}

In the experiments, we calculate the DTW values between the resource utilization curves of all machines and the selected standard curves.
 For instance, we select 5 curves as the standard curves, which are the standard curve \emph{16, 19, 23, 28, 36}\footnote{Here, standard curves \emph{16} represents the resource utilization curve on machine \emph{16}.}.  Then we plot the sorted DTW values that between all machines and the standard curves in Figure~\ref{dtw-machine}.
  We can see that, the standard curve \emph{23} is slightly different from  other standard curves, so we think that the resource utilization curve on machine 23 may be not suitable as the standard curve.

In addition, based on the average DTW value of standard curve \emph{16, 19, 28} and \emph{36}, we calculate the standard value of DTW is about 1.72,
and plot the proportion of different DTW ranges in Figure~\ref{dtw-range}.
There are 478 nodes whose DTW value is in the range of 1 to 2, which has the largest proportion.
About 50\% of nodes have a DTW value that is greater than 2, and 7\% of nodes have a DTW value that is greater than 5.
By manually analyzing the  resource utilization curves, we find that, when the DTW value is greater than 3,
there may be a big gap between this resource curve and the selected standard curve.
Assuming that 3 is the threshold of DTW value for judging the abnormal node, and there are  46\% of the nodes that will be divided into abnormal nodes. That is, the performance of different nodes is different and volatility.

\textbf{summary}. The performance discrepancy of the machines in Alibaba's co-located workloads cluster is relatively large.

   \begin{figure*}[htbp]
   \begin{center}
   \includegraphics[height=3.5cm,width=18cm]{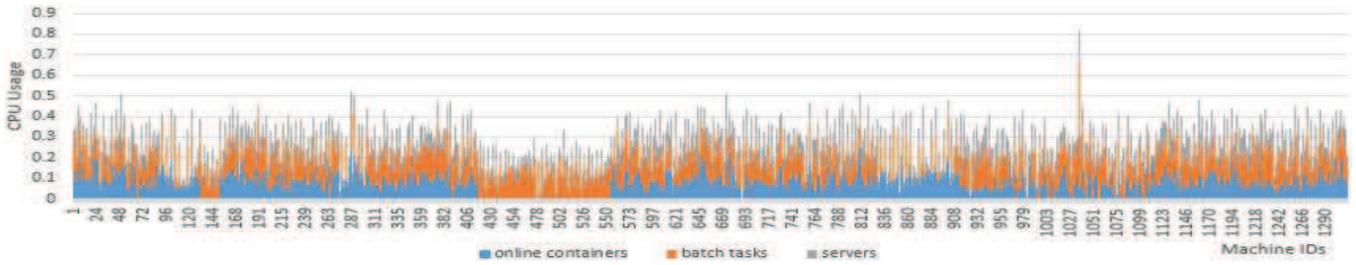}
   \end{center}
    \caption{The CPU utilization of online containers, batch tasks and servers.}
    \label{cpu_usage}
   \end{figure*}

  \begin{figure*}[htbp]
   \begin{center}
   \includegraphics[height=3.5cm,width=18cm]{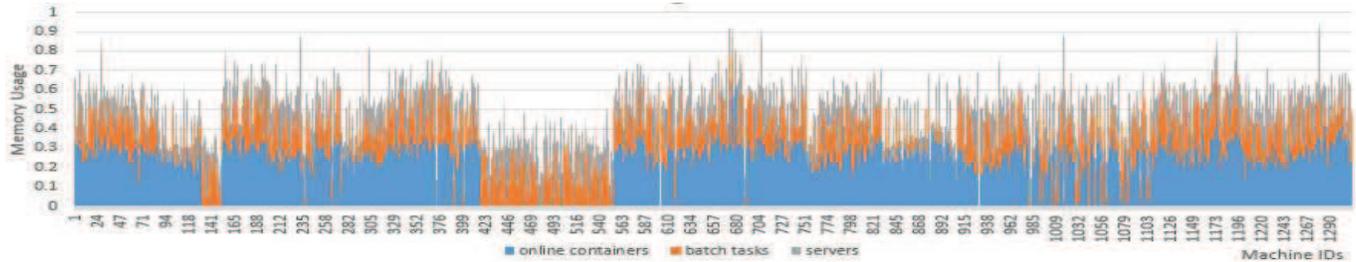}
   \end{center}
    \caption{The memory utilization of online containers, batch tasks and servers.}
    \label{mem_usage}
   \end{figure*}

\section{ Co-located Workloads Characteristics}\label{co-located}

\subsection{Resource Utilization  of Co-located Workloads}


The CPU usage and memory usage of online containers, batch tasks and servers are shown in Figure~\ref{cpu_usage} and Figure~\ref{mem_usage}.
From these two figures, we see that the resource utilization of server is slightly higher than the sum of online containers and batch tasks' resource utilization. It is  no doubt that the OS system will take up some resources.
We also observe that there are some spikes in these figures, which implies that some machines may have a sudden high resource utilization at a certain time.
From the range of machine \emph{132} to \emph{151}, machine \emph{418} to \emph{553}, they are lack of online containers' resource utilization, which implies that these machine regions are hosting batch jobs only.

\begin{figure}[tbp]
\centering
\subfigure[CPU usage.]{
\begin{minipage}[t]{0.5\linewidth}
\centering
\includegraphics[width=1.7in]{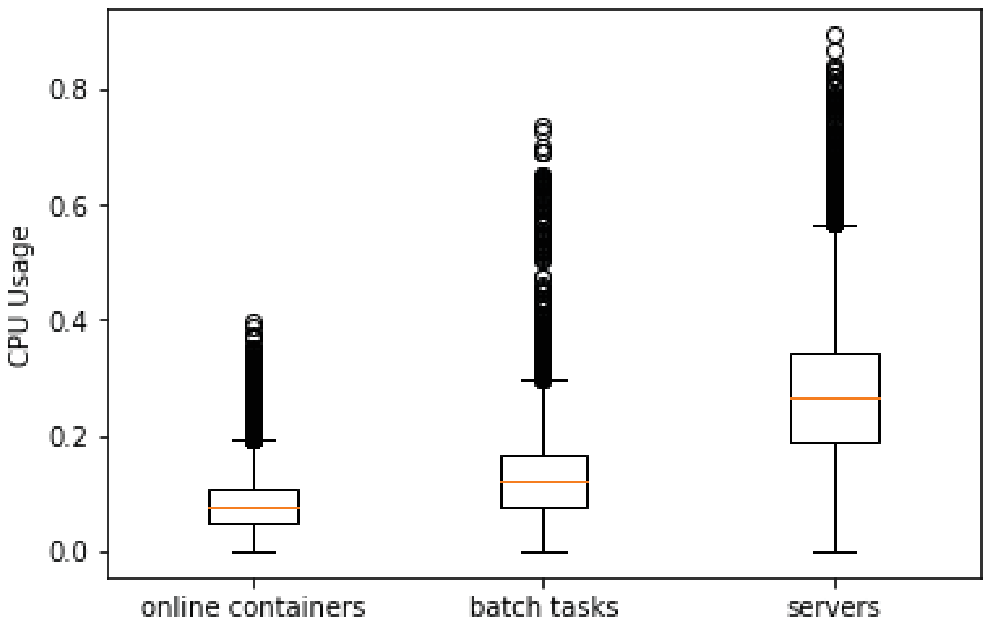}
\end{minipage}%
}%
\subfigure[Memory usage.]{
\begin{minipage}[t]{0.5\linewidth}
\centering
\includegraphics[width=1.7in]{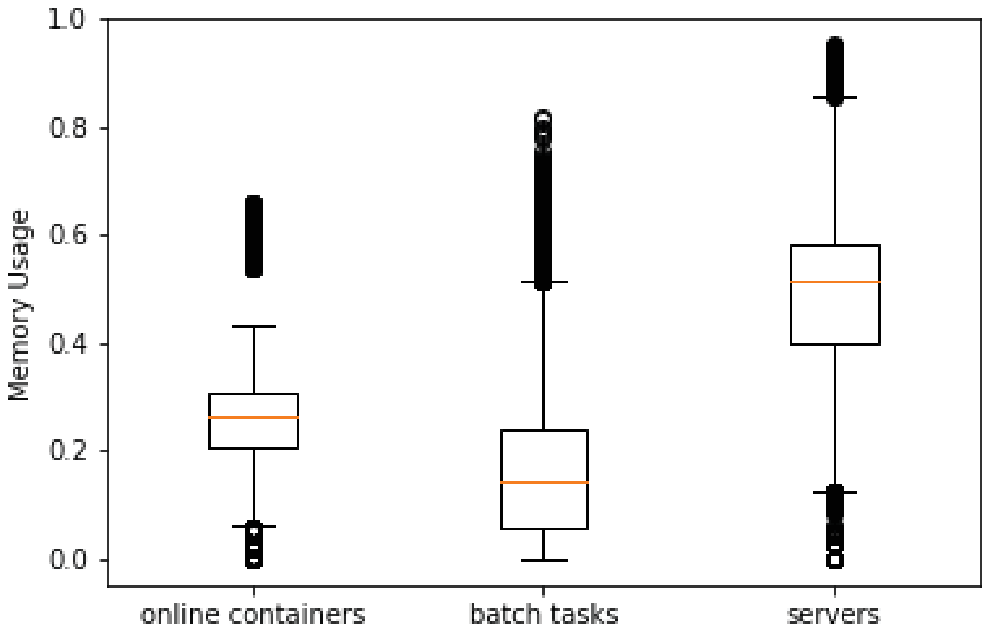}
\end{minipage}%
}%
\centering
\caption{The box-and-whisker plots that showing CPU and memory usage distribution.}
\label{box-fig}
\end{figure}

Figure~\ref{box-fig} is the box-and-whisker plots that showing CPU usage and memory usage distributions.
We observe that on the same machine, the aggregated CPU usage of  online containers  is lower than  that of batch tasks, while the aggregated memory usage of  online containers is higher than that of batch tasks.
It implies  that the online container services (long-running jobs) are more memory-demanding.


\begin{figure}[tbp]
\centering
\subfigure[CPU usage.]{
\begin{minipage}[t]{0.5\linewidth}
\centering
\includegraphics[width=1.7in]{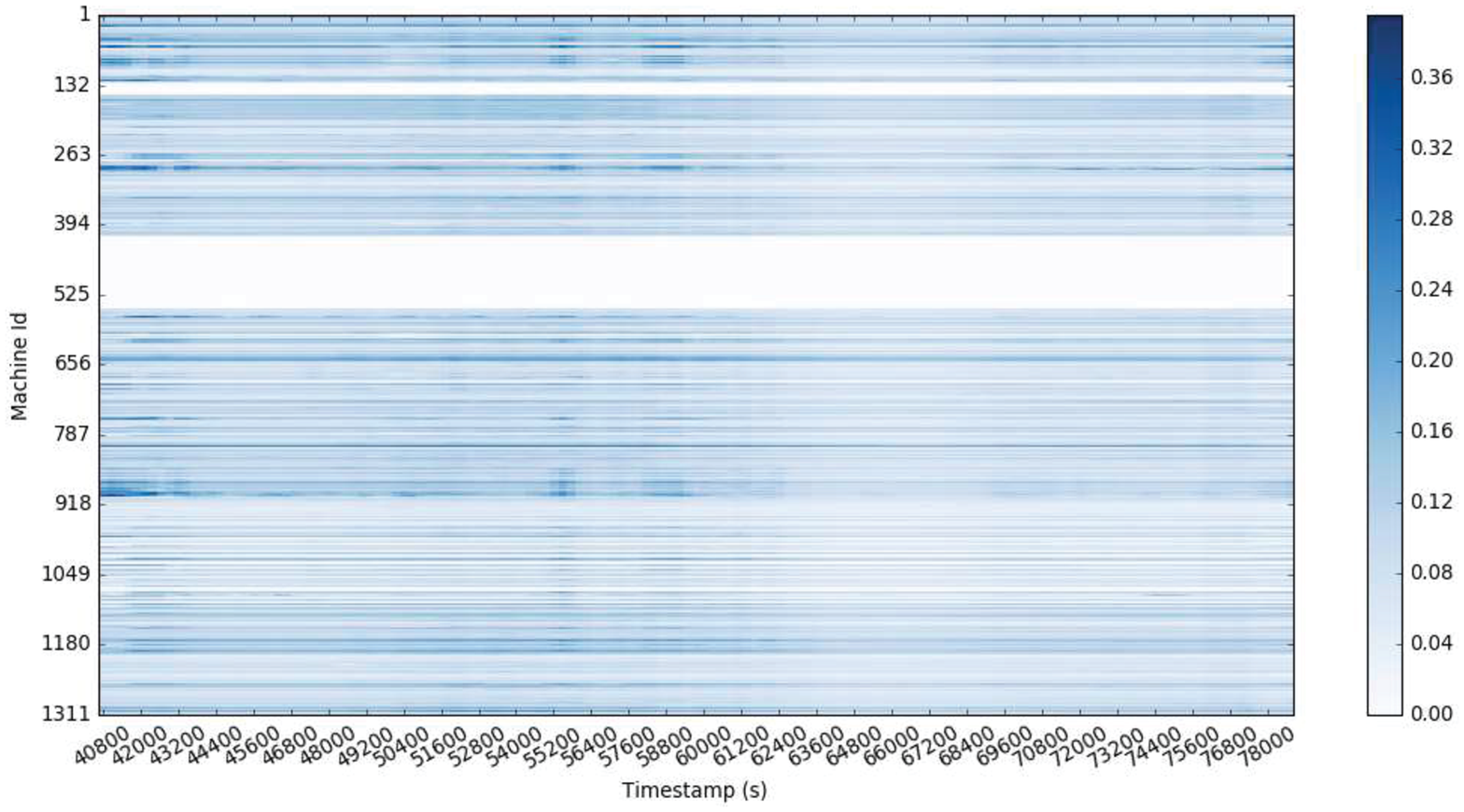}
\end{minipage}%
}%
\subfigure[Memory usage.]{
\begin{minipage}[t]{0.5\linewidth}
\centering
\includegraphics[width=1.7in]{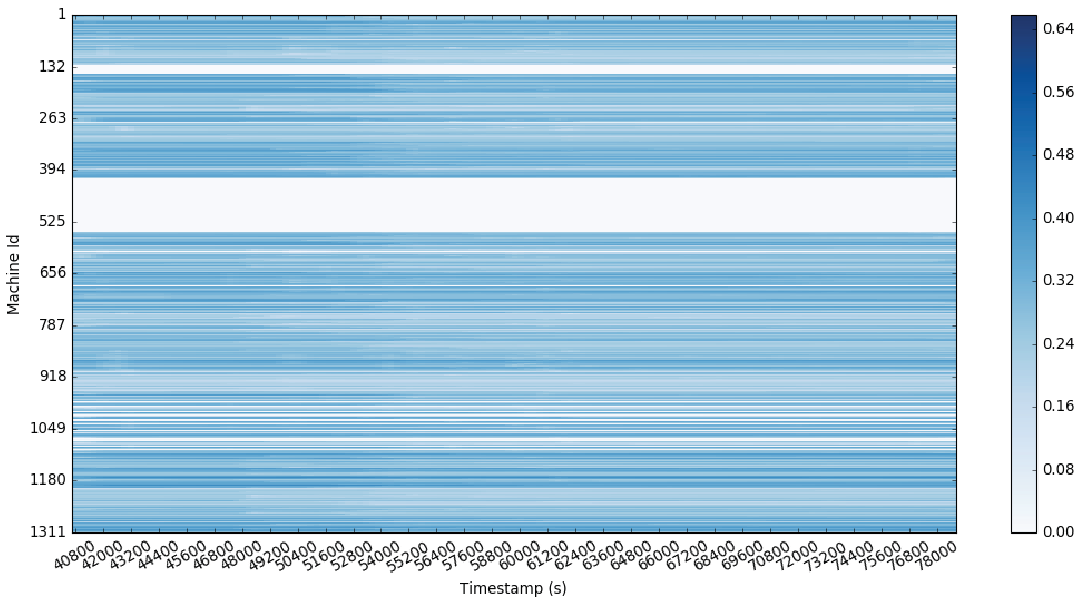}
\end{minipage}%
}%
\centering
\caption{The resource usage heatmap of online containers.}
\label{heatmap-1}
\end{figure}

\begin{figure}[tbp]
\centering
\subfigure[CPU usage.]{
\begin{minipage}[t]{0.5\linewidth}
\centering
\includegraphics[width=1.7in]{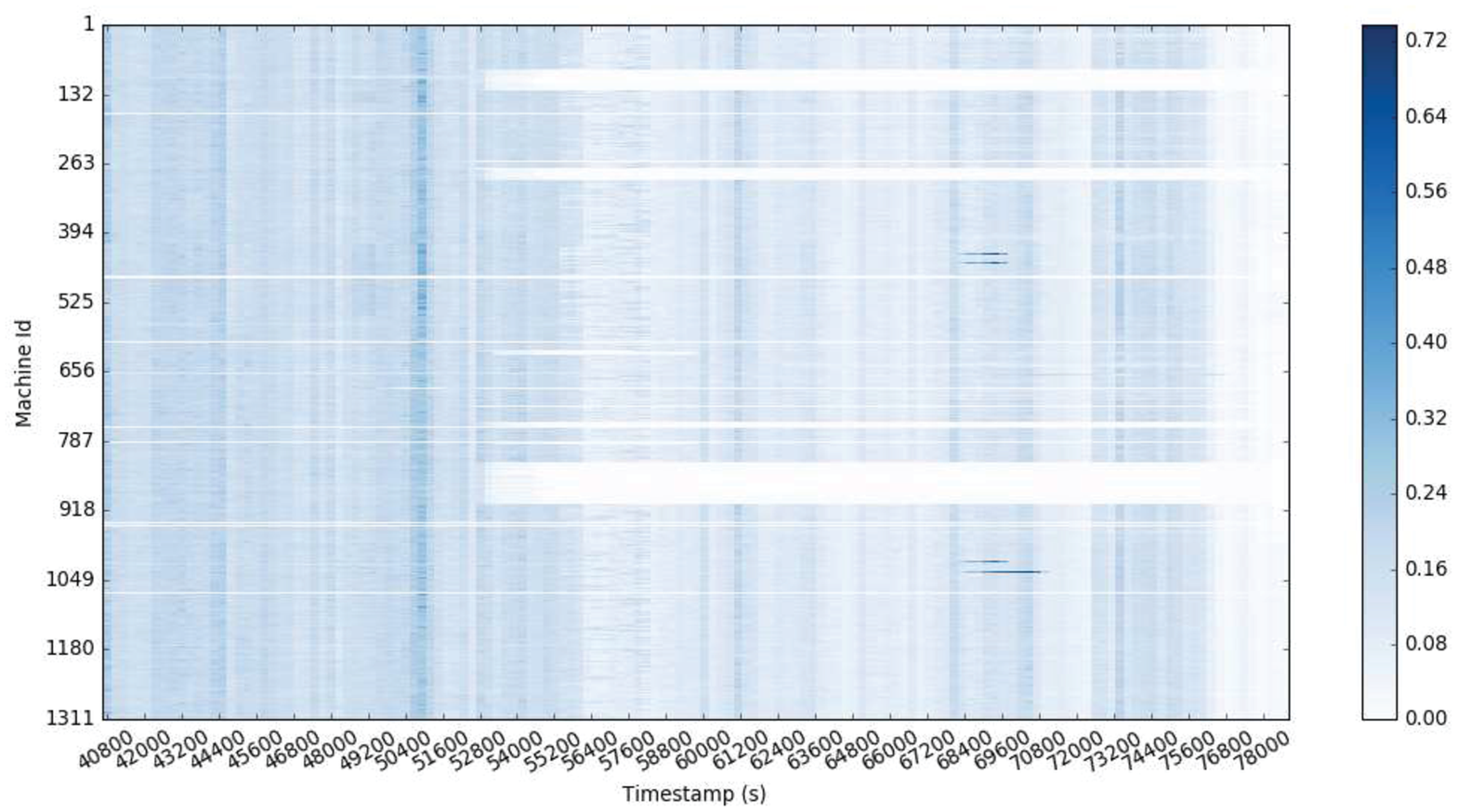}
\end{minipage}%
}%
\subfigure[Memory usage.]{
\begin{minipage}[t]{0.5\linewidth}
\centering
\includegraphics[width=1.7in]{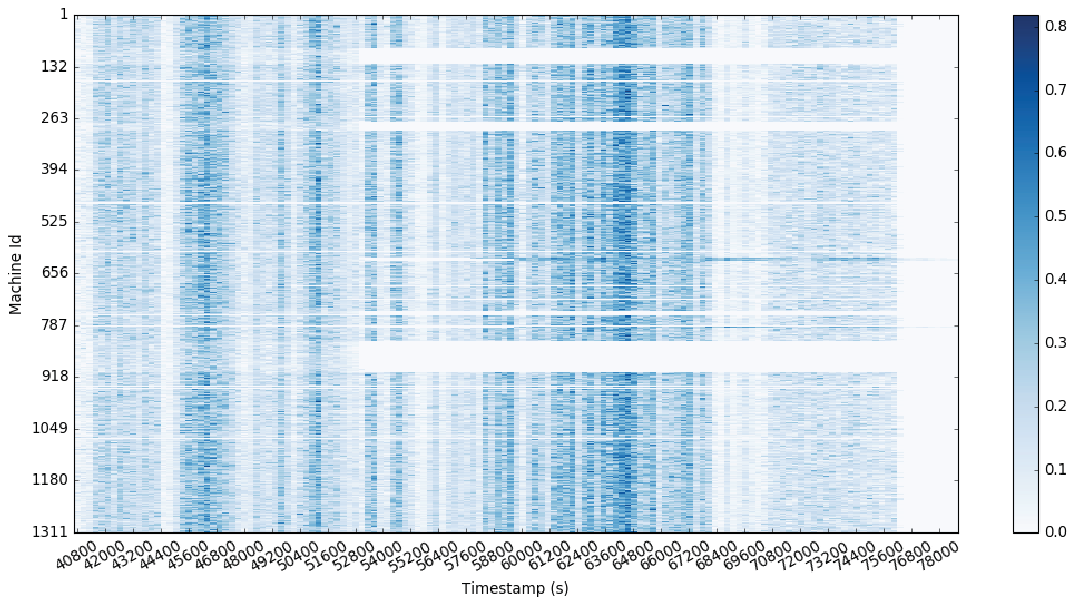}
\end{minipage}%
}%
\centering
\caption{The resource usage heatmap of batch tasks.}
\label{heatmap-2}
\end{figure}

We also plot the resource usage heatmap of online containers and batch tasks in Figure~\ref{heatmap-1} and Figure~\ref{heatmap-2}.
Figure~\ref{heatmap-1} also shows that, there are no running online containers from the range of machine \emph{132} to \emph{151}, machine \emph{418} to \emph{553}.
During the tracing interval, the resource utilization (CPU usage and memory usage) of  online containers is relatively stable.
Figure~\ref{heatmap-2} shows that, there are no running batch tasks from  52800s (14.7h) in some machine regions, such as the region of machine \emph{95} to \emph{127}, machine \emph{275} to \emph{296}, machine \emph{753} to \emph{760}, and machine \emph{830} to \emph{906}.
Since most batch tasks are short jobs, the resource utilization is not as stable as that of long-running jobs,
especially the memory usage is fluctuating.

\textbf{summary}. The online container instances and batch tasks are not running on all machines in the cluster.
 Since the online containers are the  long-running jobs with more memory-demanding, the memory usage is relatively stable; while  the memory usage of batch jobs is fluctuating for most batch tasks are short jobs.

\subsection{ Distribution Characteristics of Co-located Workloads }

\begin{figure*}[htbp]
\centering

\subfigure[Type 1.]{
\begin{minipage}[t]{0.25\linewidth}
\centering
\includegraphics[width=1.5in]{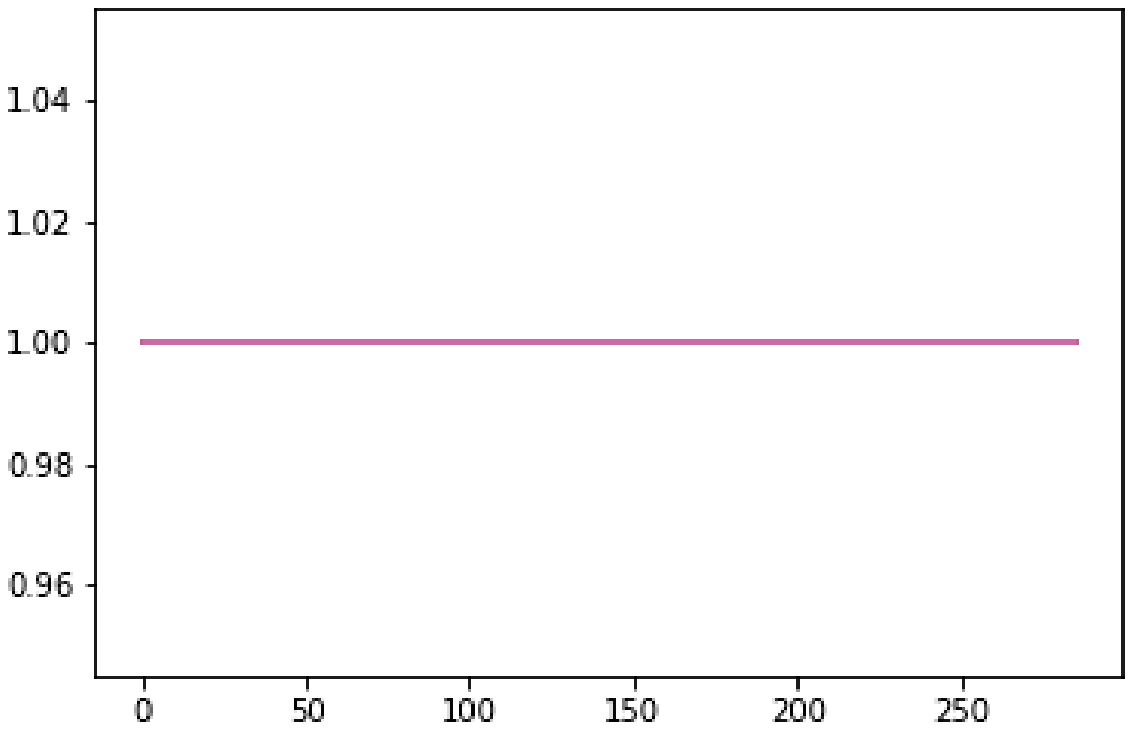}
\end{minipage}%
}%
\subfigure[Type 2.]{
\begin{minipage}[t]{0.25\linewidth}
\centering
\includegraphics[width=1.5in]{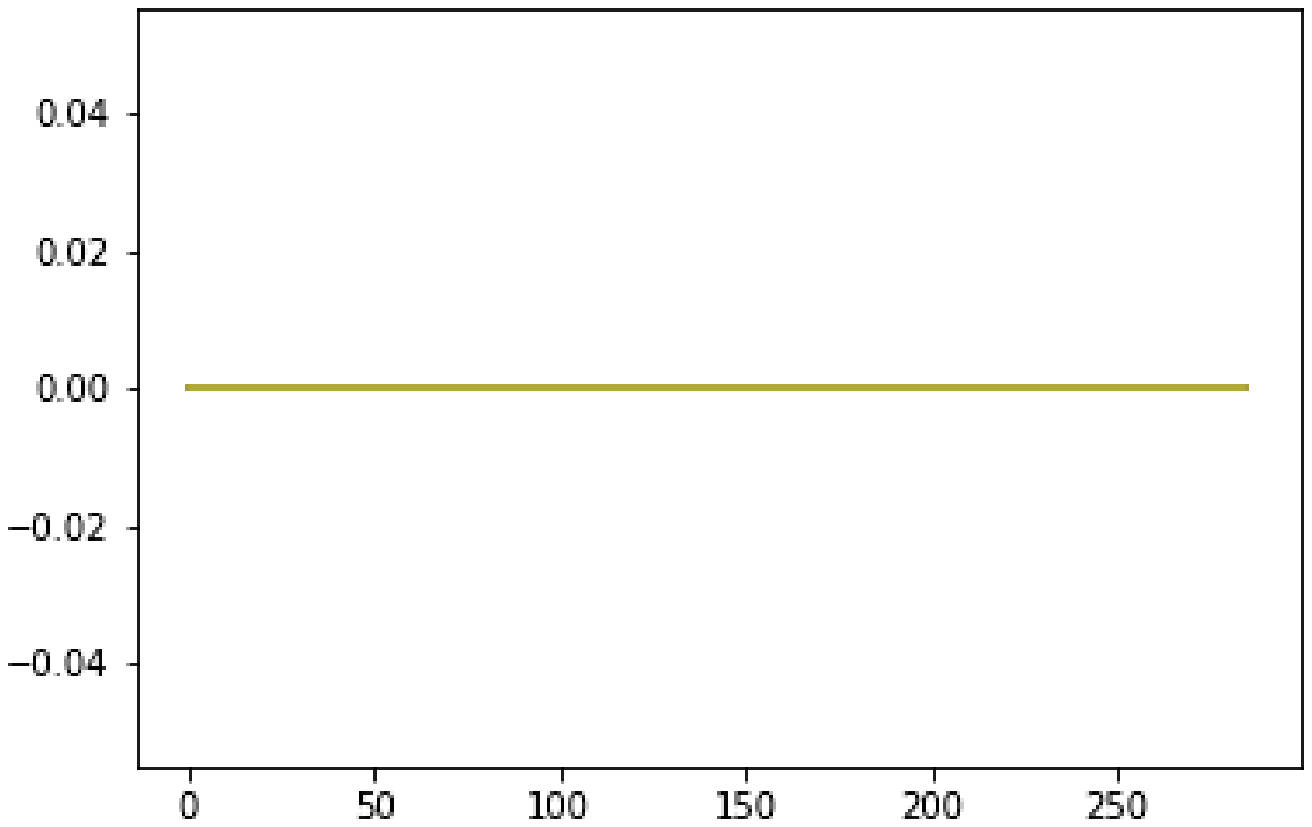}
\end{minipage}%
}%
\subfigure[Type 3.]{
\begin{minipage}[t]{0.25\linewidth}
\centering
\includegraphics[width=1.5in]{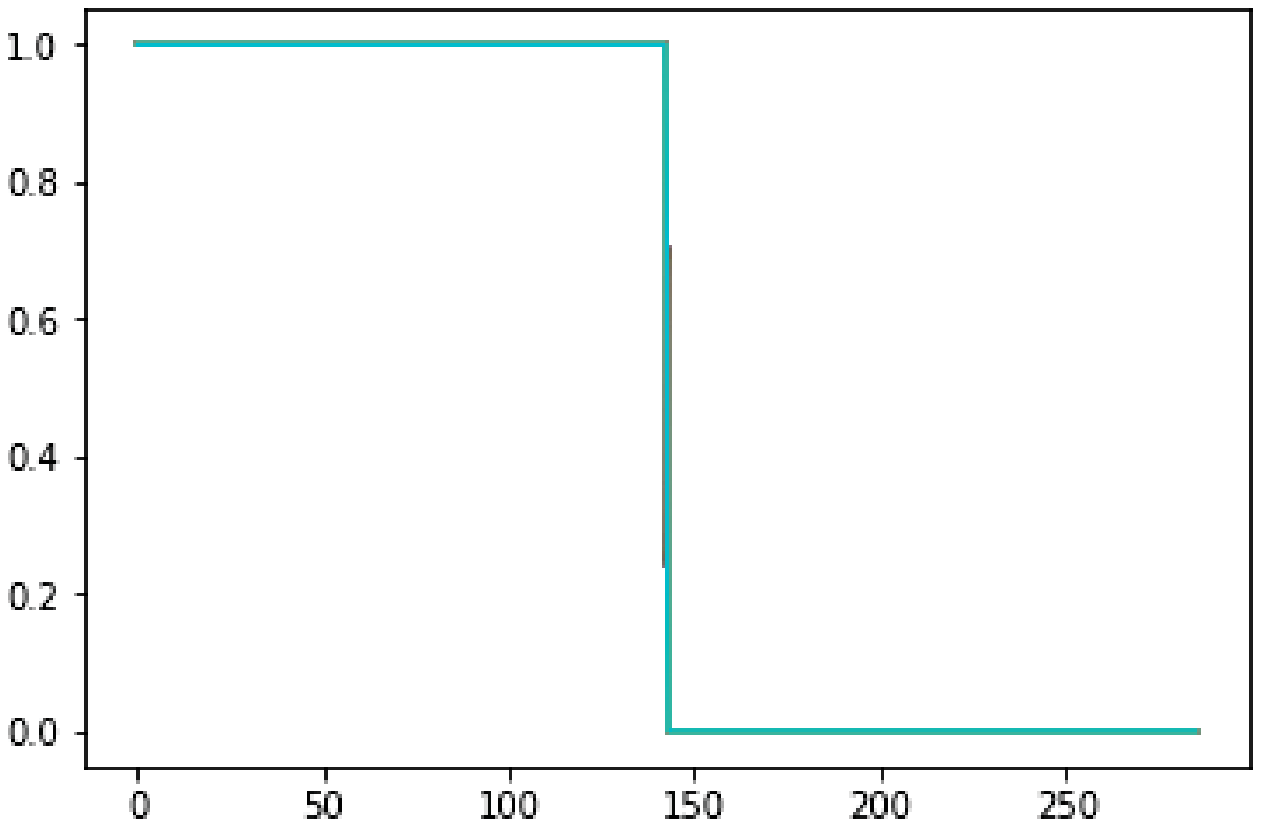}
\end{minipage}
}%
\subfigure[Type 4.]{
\begin{minipage}[t]{0.25\linewidth}
\centering
\includegraphics[width=1.5in]{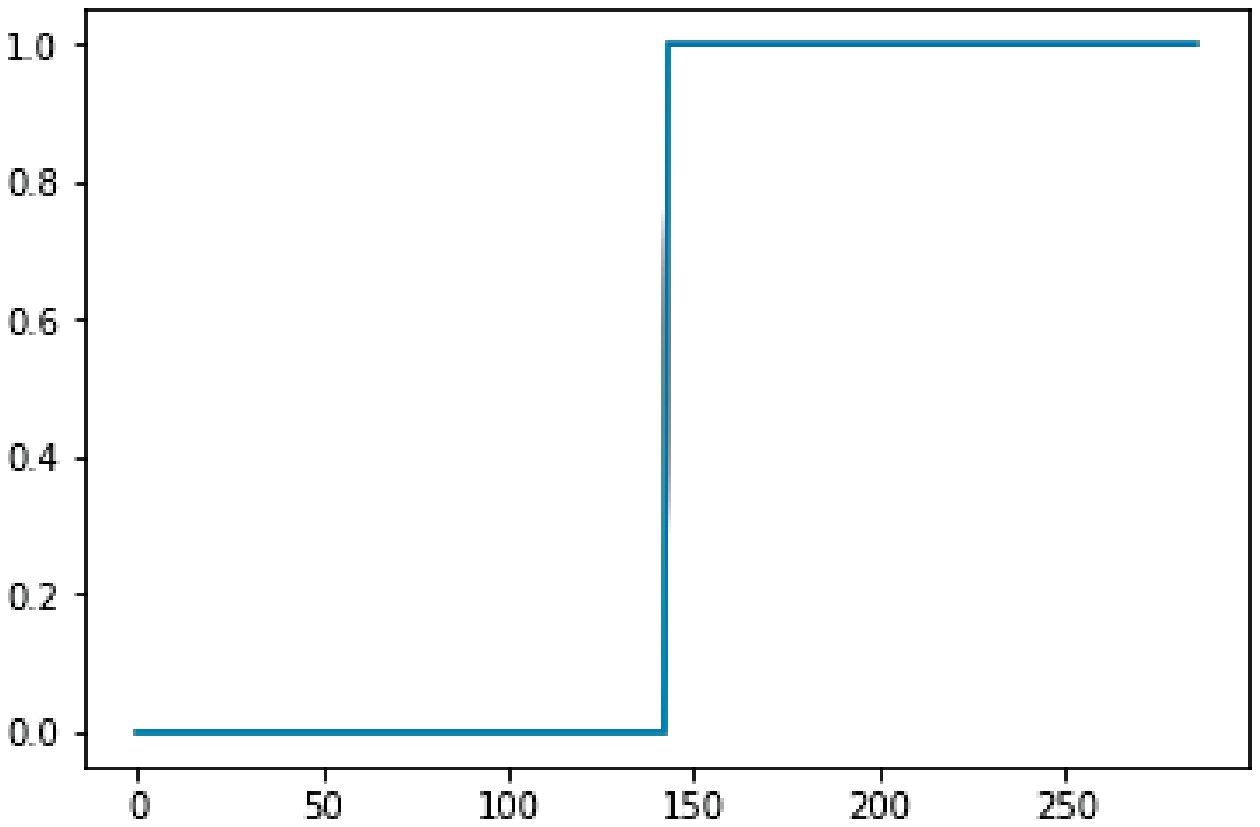}
\end{minipage}
}%

\subfigure[Type 5.]{
\begin{minipage}[t]{0.25\linewidth}
\centering
\includegraphics[width=1.5in]{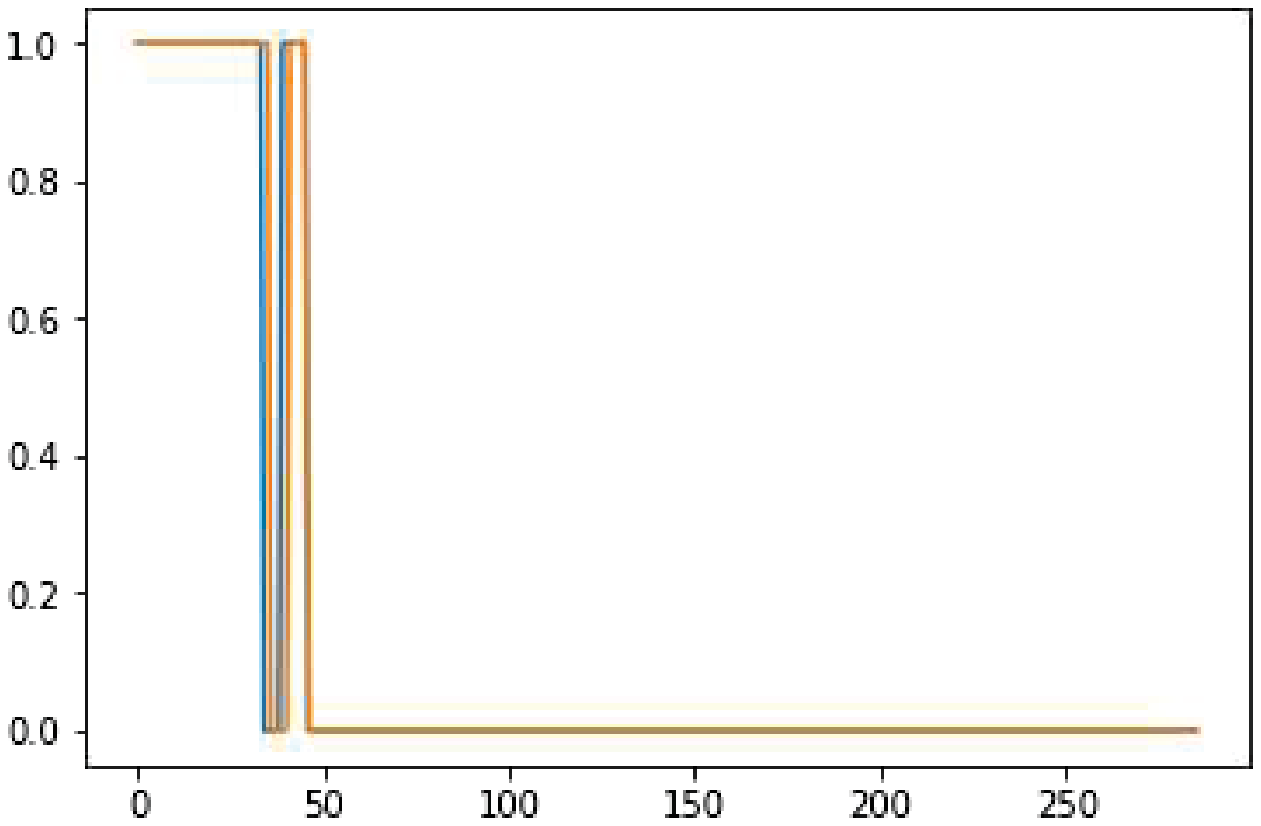}
\end{minipage}%
}%
\subfigure[Type 6.]{
\begin{minipage}[t]{0.25\linewidth}
\centering
\includegraphics[width=1.5in]{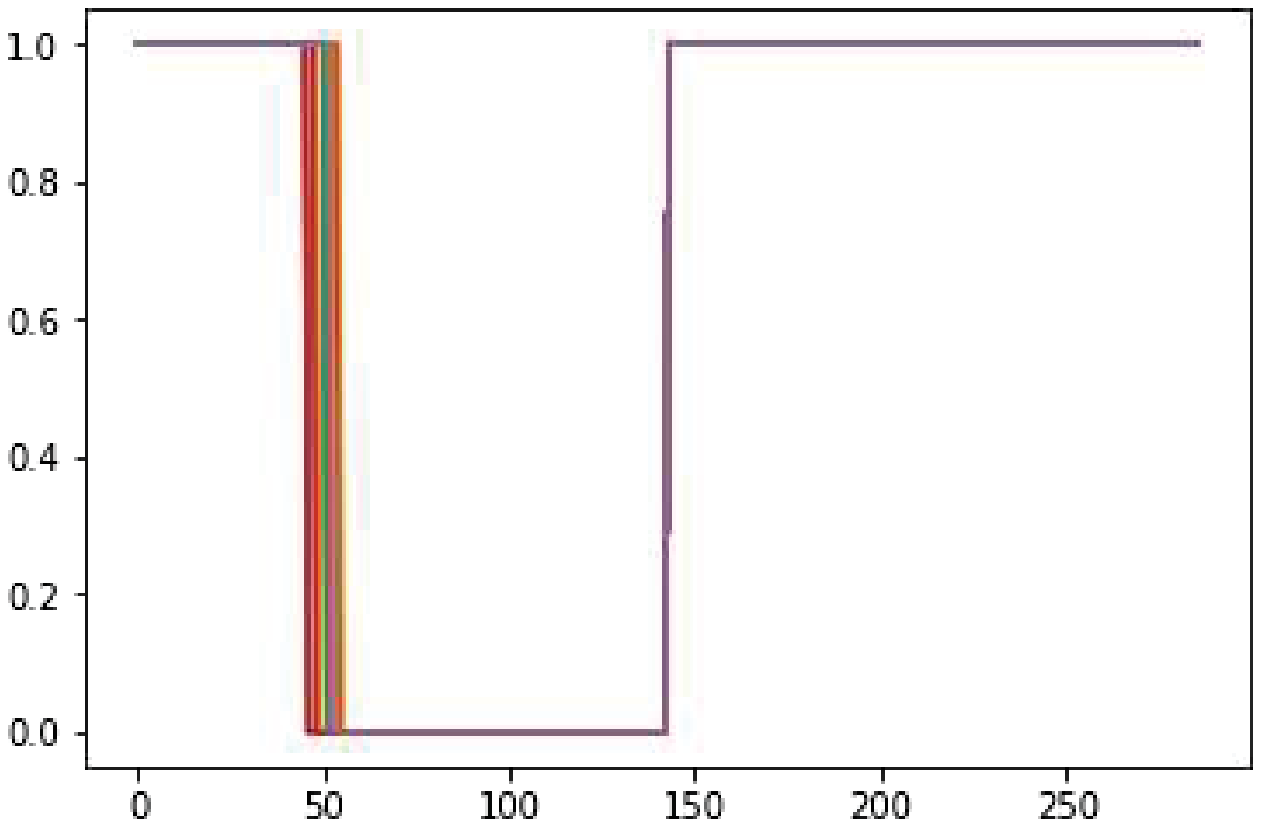}
\end{minipage}%
}%
\subfigure[Type 7.]{
\begin{minipage}[t]{0.25\linewidth}
\centering
\includegraphics[width=1.5in]{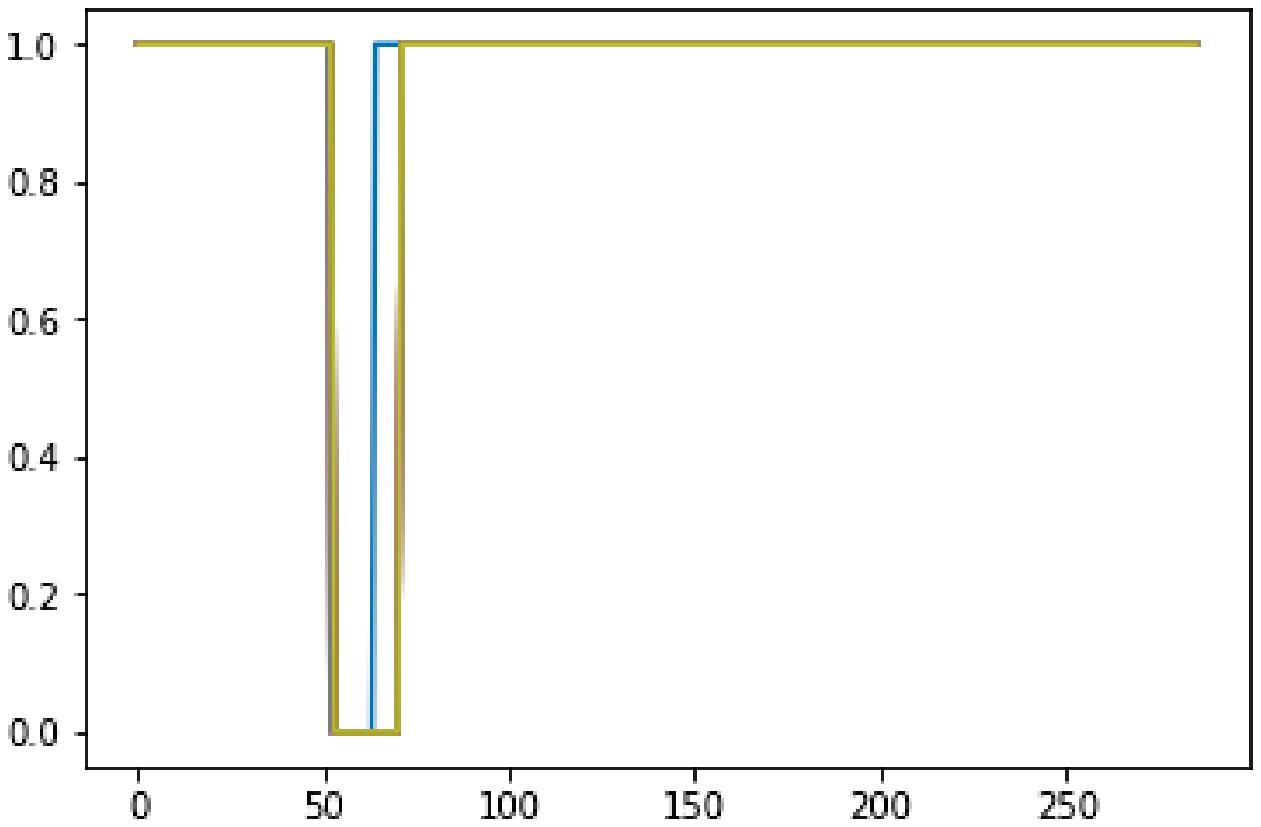}
\end{minipage}
}%
\subfigure[Type 8.]{
\begin{minipage}[t]{0.25\linewidth}
\centering
\includegraphics[width=1.5in]{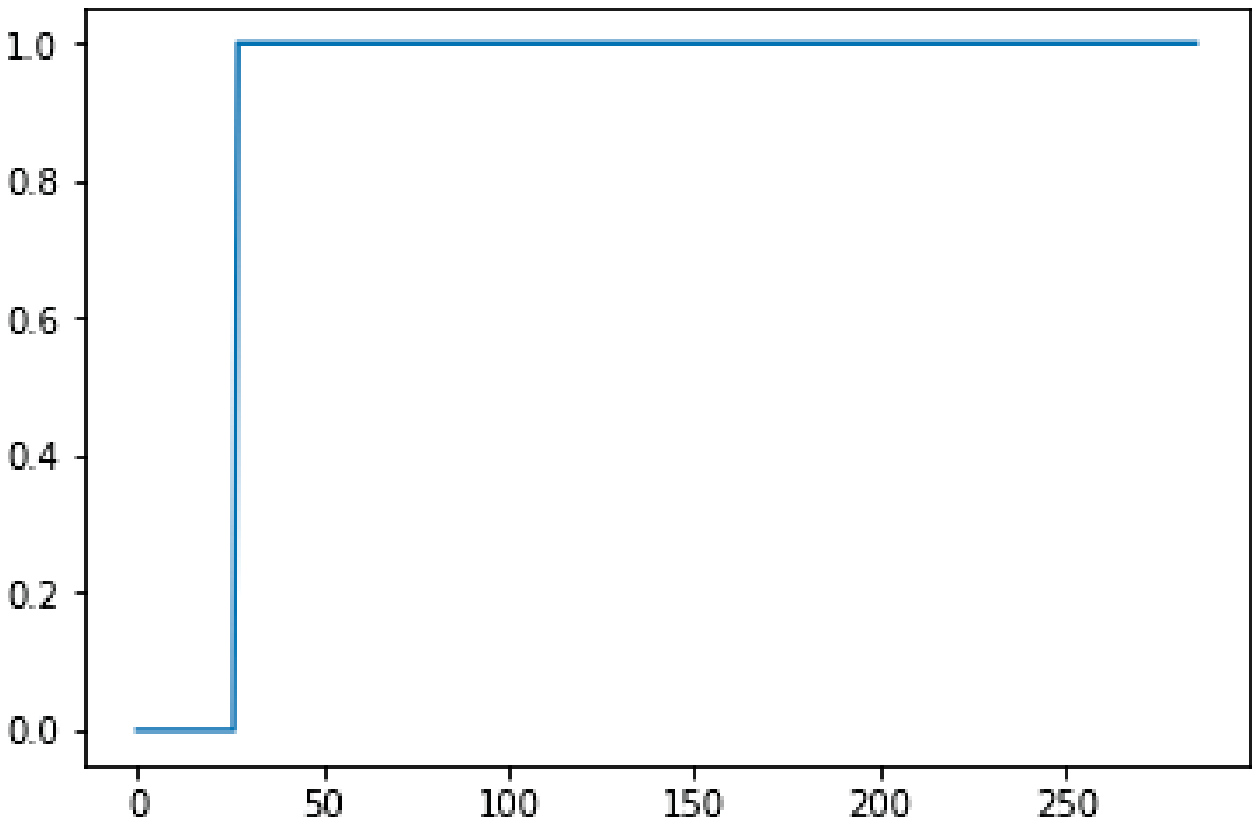}
\end{minipage}
}%
\centering
\caption{Categories of co-located workload distribution.}
\label{type}
\end{figure*}

\begin{figure}[htbp]
\centering
\subfigure[Online containers.]{
\begin{minipage}[t]{0.5\linewidth}
\centering
\includegraphics[height=1.15in,width=1.7in]{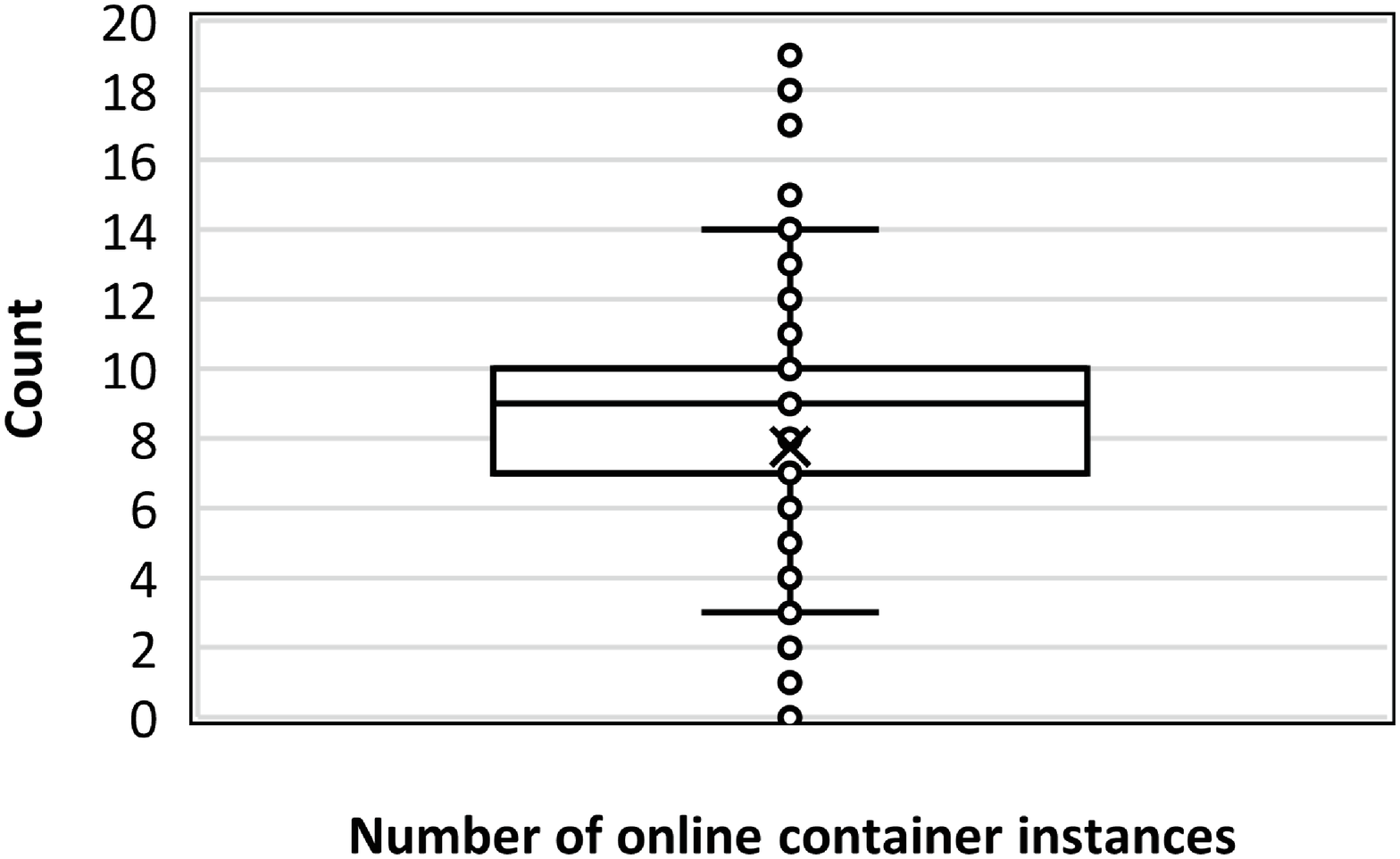}
\end{minipage}%
}%
\subfigure[Batch tasks.]{
\begin{minipage}[t]{0.5\linewidth}
\centering
\includegraphics[height=1.15in,width=1.75in]{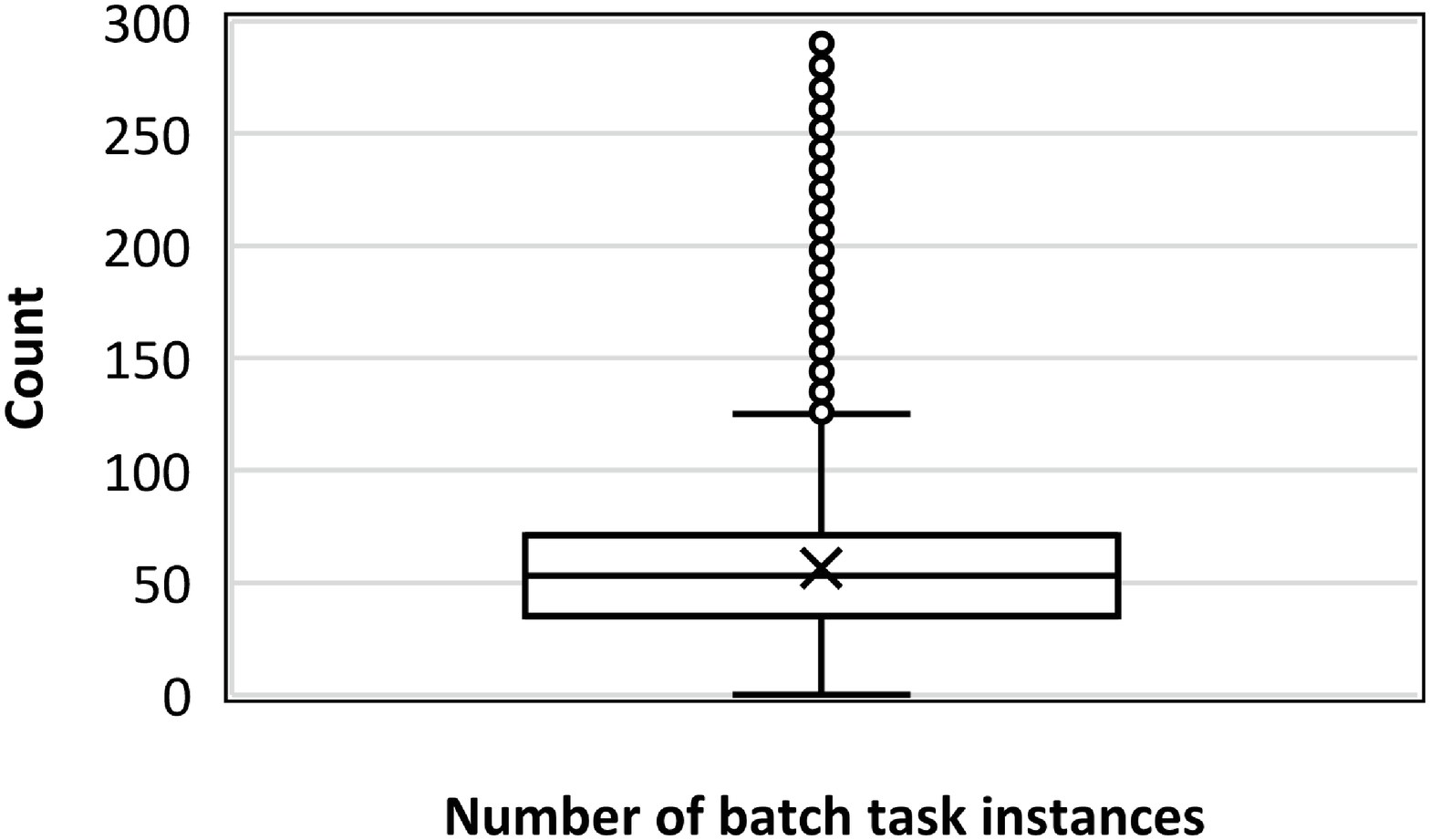}
\end{minipage}%
}%
\centering
\caption{The box-and-whisker plots about number of online container and batch tasks.}
\label{box-num}
\end{figure}

In Figure~\ref{box-num}, we give the box-and-whisker plots about the number of online container and batch tasks during every time interval.
We observe that,  most of the batch task numbers are in the range of 35 to 71, and most of the online container  numbers are in the range of 7 to 10.

Based on the number of batch tasks and online containers on machines,
we classify the distribution of the co-located workloads.
First, the non-zero values of $Num(bi)_{m,I_{x}}$ and $Num(ci)_{m,I_{x}}$ are mapped to 1, the zero values remains unchanged.
Second, for each machine, we combine all the mapped batch task numbers and container numbers to form a (143+143)-dimensional\footnote{The number of recording interval is 143.} vector.
That is, it generates a matrix of 1313*286.
At last, the Kmeans~\cite{k-means} algorithm is applied to the generated number matrix and is used for classification.
 All machines in Alibaba cluster can be classified into 8 \emph{workload distribution categories}, and the machine number that belonging to these 8 categories is shown in Table ~\ref{type-count}.
In detail, the 8 workload distribution categories include:

\begin{itemize}
\item \textbf{Type 1}: The online containers and batch tasks are always co-located running on machines, which is shown in Figure~\ref{type} (a).

\item \textbf{Type 2}: No running workloads on machines, which is shown in Figure~\ref{type} (b).

\item \textbf{Type 3}: Batch tasks are running only, which is shown in Figure~\ref{type} (c).  

\item \textbf{Type 4}: Online container instances are running only, which is shown in Figure~\ref{type} (d).

\item \textbf{Type 5}: Batch tasks are running only during the first few hours of tracing, which is shown in Figure~\ref{type} (e).

\item \textbf{Type 6}: The online containers and batch tasks are co-located on machines, but no batch tasks run during the latter few hours of tracing, which is shown in Figure~\ref{type} (f).

\item \textbf{Type 7}: The online containers and batch tasks are co-located on machines, but no batch tasks run during a short time of tracing, which is shown in Figure~\ref{type} (g).

\item \textbf{Type 8}: The online containers and batch tasks are co-located on machines, but no batch tasks run during  the first few hours of tracing, which is shown in Figure~\ref{type} (h).
\end{itemize}

\begin{table}[htpb]
\renewcommand{\arraystretch}{1.3}
\scriptsize
\centering
\caption{The machine number of workload distribution categories.}
\begin{tabular}{|p{0.7cm}|p{0.7cm}|p{0.7cm}|p{0.7cm}|p{0.7cm}|p{0.7cm}|p{0.7cm}|p{0.7cm}|}
  \hline
Type 1 & Type 2 & Type 3  &   Type 4  & Type 5  & Type 6  & Type 7  & Type  8 \\ \hline
956 & 9 & 170 &   11 &  2  & 155 & 9  & 1 \\ \hline
\end{tabular}
\label{type-count}
\end{table}

 \begin{figure}
   \begin{center}
   \includegraphics[scale=0.7]{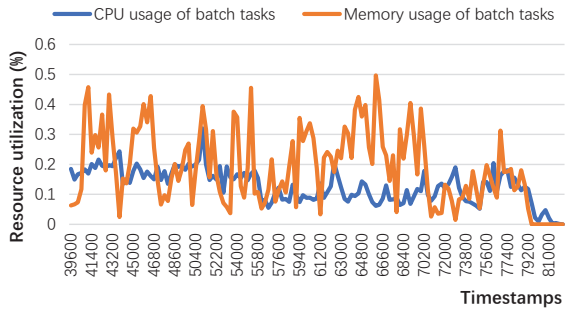}
   \end{center}
    \caption{Resource utilization of batch tasks in machine \emph{149}.}
    \label{usage-149}
   \end{figure}

\begin{figure*}
   \begin{center}
   \includegraphics[scale=0.5]{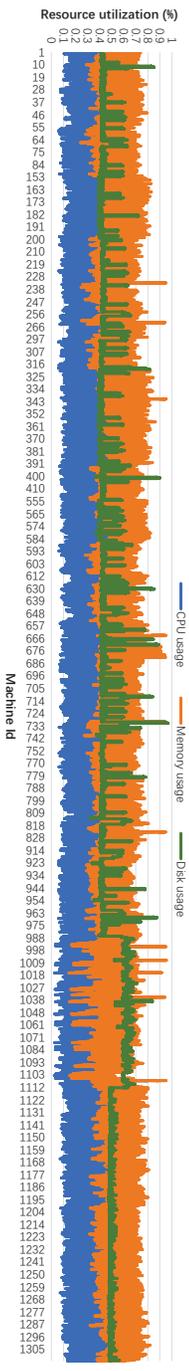}
   \end{center}
    \caption{Resource utilization of Type 1.}
    \label{type1-usage}
   \end{figure*}

\begin{figure*}[htbp]
\begin{minipage}[t]{0.33\linewidth}
\centering
\includegraphics[height=3cm,width=5cm]{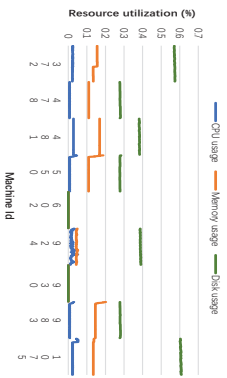}
\caption{Resource utilization of Type 2.}
 \label{type2-usage}
\end{minipage}%
\begin{minipage}[t]{0.33\linewidth}
\centering
\includegraphics[height=3cm,width=5cm]{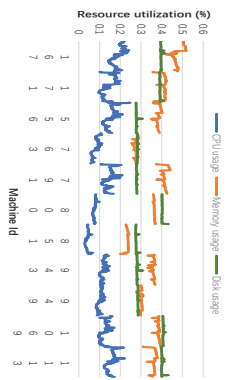}
\caption{Resource utilization of Type 4.}
 \label{type4-usage}
\end{minipage}
\begin{minipage}[t]{0.33\linewidth}
\centering
\includegraphics[height=3cm,width=5cm]{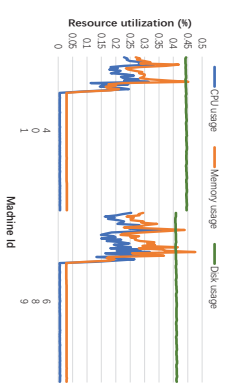}
\caption{Resource utilization of Type 5.}
\label{type5-usage}
\end{minipage}%
\end{figure*}

\begin{figure*}
   \begin{center}
   \includegraphics[scale=0.5]{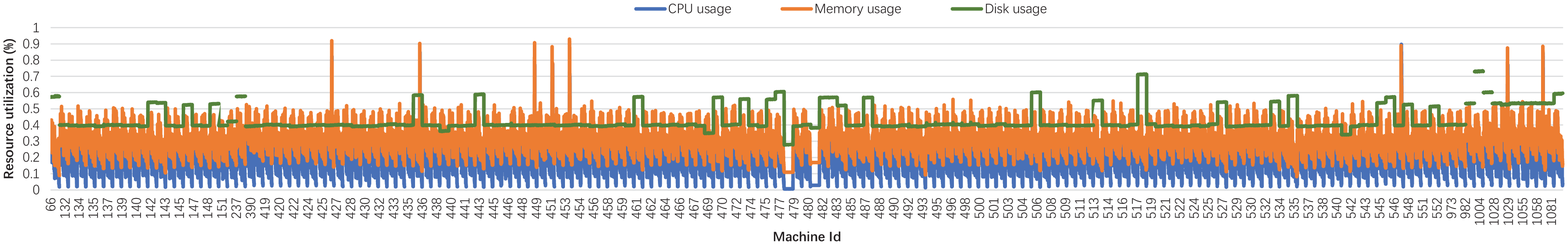}
   \end{center}
    \caption{Resource utilization of Type 3.}
    \label{type3-usage}
   \end{figure*}

\begin{figure*}
   \begin{center}
   \includegraphics[scale=0.5]{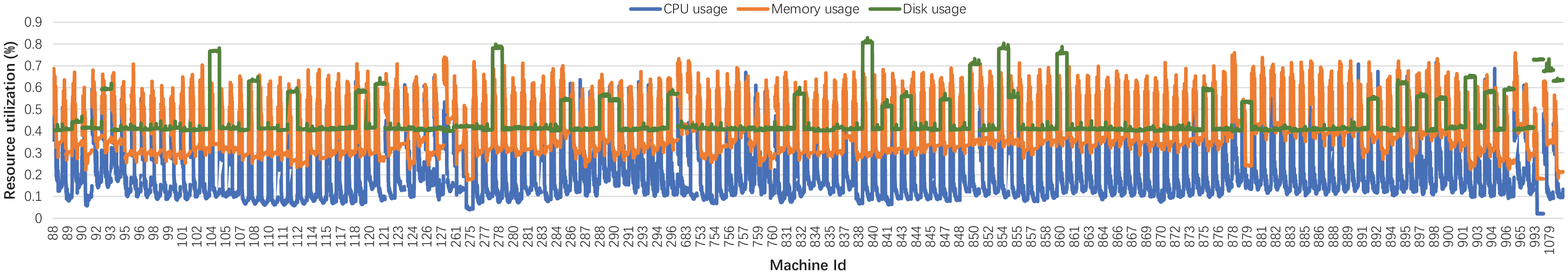}
   \end{center}
    \caption{Resource utilization of Type 6.}
    \label{type6-usage}
   \end{figure*}

\begin{figure*}[htbp]
\begin{minipage}[t]{0.56\linewidth}
\centering
\includegraphics[height=3cm,width=8cm]{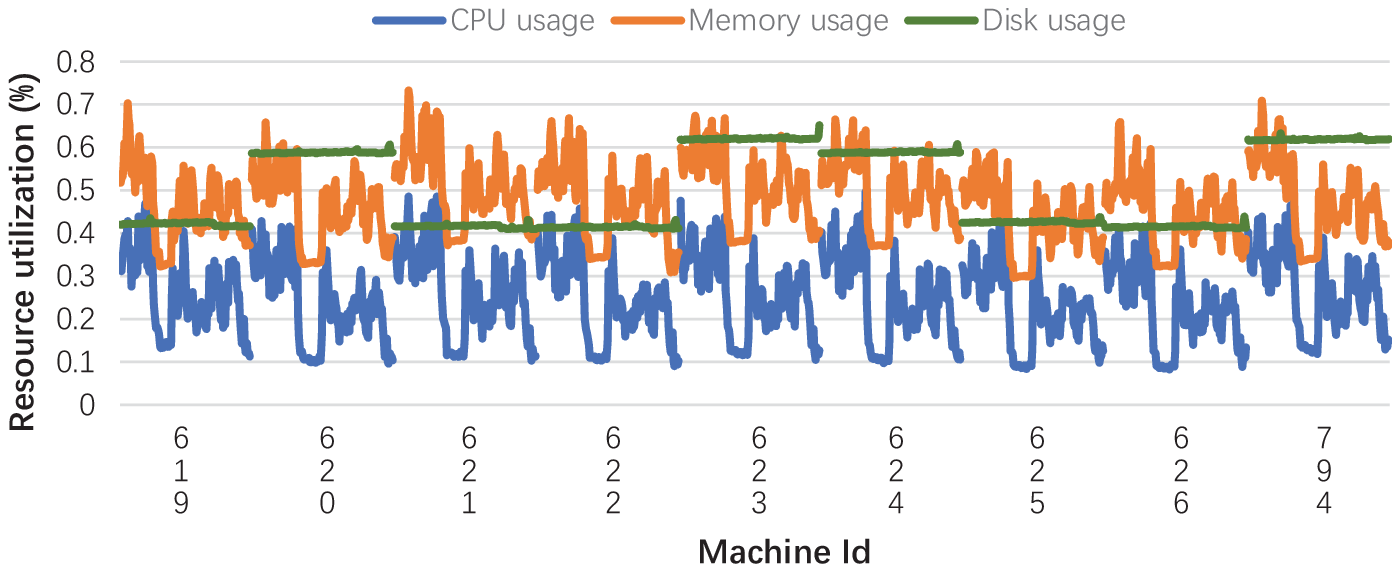}
\caption{Resource utilization of Type 7.}
\label{type7-usage}
\end{minipage}
\begin{minipage}[t]{0.4\linewidth}
\centering
\includegraphics[height=3cm,width=5cm]{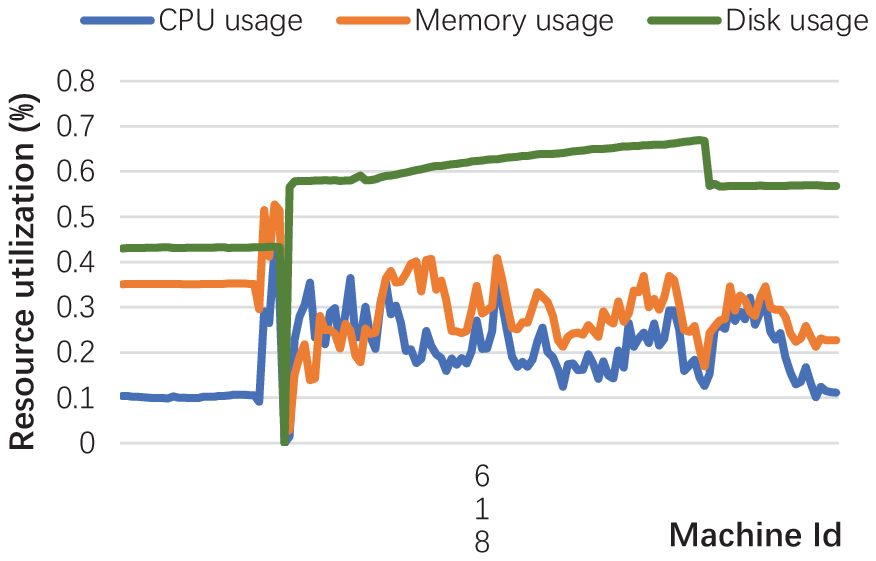}
\caption{Resource utilization of Type 8.}
\label{type8-usage}
\end{minipage}
\end{figure*}

From Table ~\ref{type-count} we see that, 72.8\% of nodes have the co-located workloads, and they belong to \textbf{Type 1}.
The resource utilization curves on these nodes are shown in Figure~\ref{type1-usage}:
The CPU usage, memory usage and disk usage are in the approximate  range of 20\%-30\%, 50\%-60\%, and 40\%-60\%, respectively.
There are no running workloads  on 9 nodes that belonging to \textbf{Type 2}, and the machine ids are \emph{372, 478, 481, 550, 602, 924, 930, 983, 1075}.
The resource utilization curves on these nodes are shown in Figure~\ref{type2-usage}:
the CPU usage is very low on these nodes, which is about 1\%;
and the average memory usage is 9.6\% and the average disk usage is high, which is 30.92\%.
In addition, we also find that the machine \emph{149} that lacking recorded resources is not belonging to \textbf{Type 2},
because there are some batch jobs are running on the machine \emph{149} actually.
The resource utilization of batch tasks on machine \emph{149} is shown in Figure~\ref{usage-149}.
There are 170 nodes that belonging to \textbf{Type 3}, which including: \emph{66, 132-151,  237, 265, 390, 418-549, 551-553, 973, 982, 987, 1004, 1008, 1028, 1029, 1043, 1055, 1057, 1058, 1081, 1083}, whose average CPU usage, memory usage and  disk usage are 17.44\%, 29.55\%, 43.32\%, respectively.
From Figure~\ref{type3-usage} we observe that, excluding some peak memory usages, the memory usage and CPU usage curves of different machines are similar, and the disk usage on the same machine is relatively stable with no changing.
There are 11 nodes that only have online containers (belonging to \textbf{Type 4}), which including: \emph{161, 171, 556, 763, 791, 800, 851, 943, 949, 1069, 1113}.  On these machines, the patterns of resource utilization are not obvious.
The average CPU usage, memory usage and  disk usage are respectively 12.06\%, 36.2\%, 33.46\%, and the memory usage are higher because of  the online container services requiring more memory.
There are just 2 nodes that belonging to \textbf{Type 5}, which including: \emph{401, 689}.
From Figure~\ref{type5-usage} we see that, the average CPU usage and memory usage are 22\% and 28.5\% during the batch task execution period, while the disk usage is almost  stable at 42\%.
There are 150 nodes that belonging to \textbf{Type 6}, which including: \emph{88-127,  275-296, 683, 723, 753-760, 830-850, 852-906,  965, 986, 993, 1079, 1096}, whose average CPU usage, memory usage and  disk usage are 21.29\%, 39.88\%, 45.31\%.
And there are 9 nodes that belonging to \textbf{Type 7}, which including: \emph{619-626, 794},  whose average CPU usage, memory usage and  disk usage are 24.74\%, 47.43\%, 50.04\%.
From Figure~\ref{type6-usage} and \ref{type7-usage} we also observe that, the CPU usage and memory usage curves of different machines are similar, and the disk usage on the same machine is relatively stable with no changing, too.
There is only one machine ~\emph{618} that belonging to \textbf{Type 8}, and the average CPU usage, memory usage and  disk usage are 19.58\%, 29.66\%, 56.58\%, which is shown in Figure~\ref{type8-usage}.

\textbf{Summary}. Based on the the number of batch tasks and online containers (called as \emph{co-located workload distributions}),
the machines of Alibaba's cluster can be classified into 8 workload distribution categories. In addition, for most categories,
the CPU usage and memory usage of machines that belonging to the same category have similar patterns, while disk usage of different nodes may vary greatly.


\section{Anomaly  Analysis}

Through the node similarity based on DTW value (Section~\ref{similarity}) or the co-located workloads characteristics (Section~\ref{co-located}), we could discover the abnormal nodes from different perspectives.
However, when the standard curve is not selected well, it may have a bad impact on the abnormal detection results by using DTW method.
And anomaly analysis based on the co-located workloads characteristics is a qualitative analysis.
So based on the generated associated data, we  utilize Isolation Forest (iForest)~\cite{Isolation_forests} to filter out the outliers,
and then analyze the anomalies based on co-located workload characteristics and machine states.

\subsection{Anomaly Discovery based on iForest}

\begin{figure*}
   \begin{center}
   \includegraphics[scale=0.7]{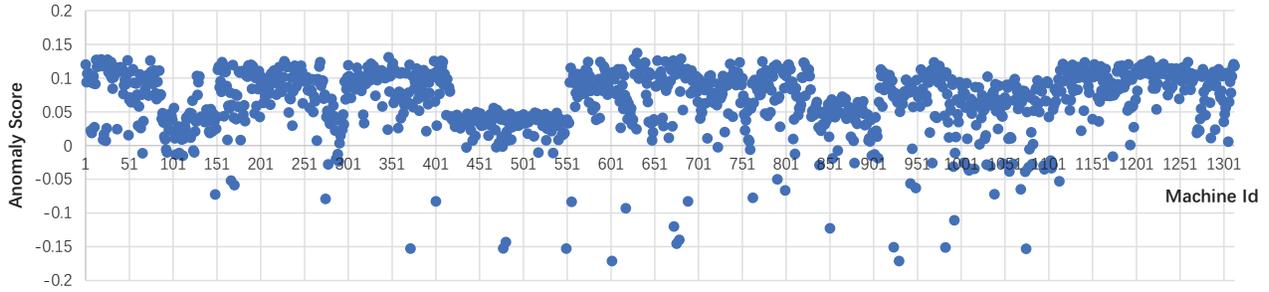}
   \end{center}
    \caption{The anomaly score.}
    \label{Anomaly-Score}
   \end{figure*}

We choose 5 dimensions $CpuUsage_{m,I_{x}}$, $MemUsage_{m,I_{x}}$, $DiskUsage_{m,I_{x}}$, $Num(bi)_{m,I_{x}}$ and $Num(ci)_{m,I_{x}}$) to build the machine-resources matrix. Then we apply the Isolation Forest (iForest)~\cite{Isolation_forests}  algorithm to this machine-resources matrix, and output the anomaly scores.
The iForest~\cite{Isolation_forests} is a fast anomaly detection method that based on ensemble, which has linear time complexity and high precision.
If one machine's anomaly score is smaller, the probability that it is an abnormal node is higher.
The distribution of machines' anomaly score is shown in  Figure~\ref{Anomaly-Score}.
Some machines have  anomaly scores that are less than 0, and the number is 81.
We also list the top 25 abnormal nodes in Table~\ref{sort-anomaly-score}.

\begin{table}[htpb]
\renewcommand{\arraystretch}{1.3}
\footnotesize
\centering
\caption{The top 25 abnormal nodes.}
\begin{tabular}{|p{0.4cm}|p{0.5cm}|p{1.8cm}|p{1.1cm}|p{2.7cm}|}
  \hline
 Top & $m$ & Anomaly score   &   Categories  & Causes \\ \hline
1 & \emph{602} &  -0.170951862 & Type 2 &  No workloads  \\
2 & \emph{930} &  -0.170951862 &   Type 2 & Frequent softerror  \\
3 & \emph{1075} & -0.152726265  &  Type 2 & Frequent softerror   \\
4 & \emph{550} & -0.152597894 &  Type 2  &   No workloads  \\
5 & \emph{372} &  -0.152429127    &  Type 2  &  Frequent softerror \\
6 & \emph{478} &  -0.152156505  & Type 2 & No workloads   \\
7 & \emph{983} &   -0.150834127 & Type 2 & No workloads  \\
8 & \emph{924} &  -0.150572048 &  Type 2 &  No workloads  \\
9 & \emph{676}  &   -0.14511185 & Type 1  &  Heavier online services \\
10 & \emph{481} &  -0.142787057 &  Type 2  & No workload \\
11 & \emph{679}	& -0.139451001 &  Type 1   & Heavier online services \\
12 & \emph{851} & 	-0.122341159 &  Type 4  & No batch jobs \\
13 & \emph{673} &	-0.119792183 &   Type 1  & Heavier online services  \\
14 & \emph{993} &	-0.110451407 & Type 6   & Unbalanced batch tasks \\
15 & \emph{618}	& -0.092764675 &  Type 8  &  Unbalanced batch tasks \\
16 & \emph{556}	& -0.083327088 &   Type 4 & No batch jobs \\
17 & \emph{689}	& -0.082675027 & Type 5  & Softerror, unbalanced workloads  \\
18 & \emph{401}	& -0.082649176 & Type 5    & Softerror, unbalanced workloads  \\
19 & \emph{275}	& -0.078791916 &  Type 6  & Unbalanced batch tasks \\
20 & \emph{763}	& -0.077354718 & Type 4   & No batch jobs \\
21 & \emph{149}	& -0.072409203 & Type 3   & No online services \\
22 & \emph{1039} &	-0.072036834 & Type 1   & Unbalanced workloads with lighter online services  \\
23 & \emph{800}	& -0.066261211 &  Type 4   &  No batch jobs \\
24 & \emph{1069} &	-0.064646667 & Type 4    &  No batch jobs \\
25 & \emph{949}	& -0.062686912 & Type 4   &  No batch jobs \\
  \hline
\end{tabular}
\label{sort-anomaly-score}
\end{table}

  \begin{figure*}
   \begin{center}
   \includegraphics[scale=0.7]{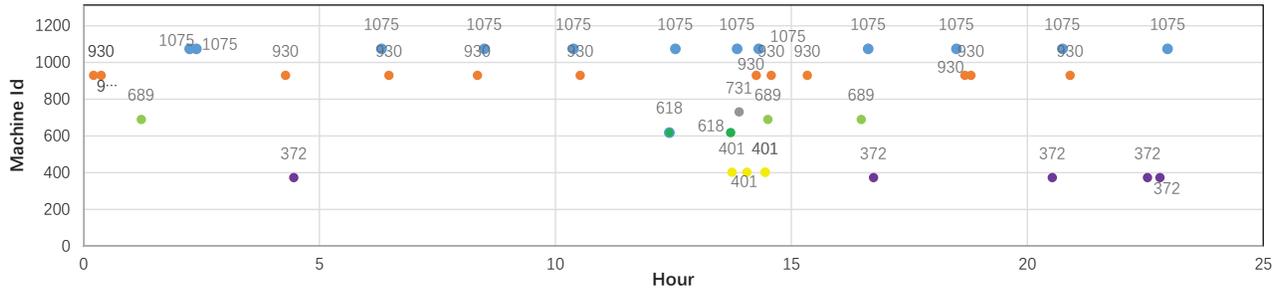}
   \end{center}
    \caption{The softerror machine.}
    \label{error_machine}
   \end{figure*}

\subsection{Cause Analysis}

We analyze the causes of  anomalies.
The one reason for anomalies is \textbf{system errors or failures}, and the softerror status of machines is shown in Figure~\ref{error_machine}:

(1) Frequent softerror can result in machines becoming unavailable, such as the machine \emph{930, 1075} and \emph{372}, with no running jobs.

(2) Due to the softerror  at a certain time, the machines  may have exceptions, which can affect the scheduling and execution of jobs.
For example, there are no online services running on the machine \emph{689} and \emph{401}, and the batch tasks are running only during the first few hours of tracing.
By checking the machine status, the machine \emph{689} has softerror at the timestamp of 50623s, 52005s and  52219s, and there is no running batch tasks  from 50400s radually;
the machine \emph{401} has softerror at the timestamp of 49854s, 50018s, 51325s and  51515s, and there is no running batch tasks  from 49800s, too.
The reason may be that, cluster management system is unable to continue scheduling and executing new jobs on these machines due to system failures.

The other reason for anomalies is \textbf{unbalanced scheduling}, which results in  workload imbalance:

(1) Obviously, due to the uneven number of online container instances and batch jobs, the imbalance of co-located workload distribution looks like an obvious reason for abnormal nodes.
For example, some non-faulty machines also belong to \textbf{Type 2}, which have no running jobs.
The possible reason is that  no tasks are assigned on these nodes based on the scheduling policies.
And on some machines, there are only batch jobs (\textbf{Type 3}) or online containers (\textbf{Type 4}), with skew workload distribution.

(2)  Skew of co-located workload resource utilization also results in  some abnormal nodes.
For instance, there are  four nodes that are belonging to \textbf{Type 1}.
And the machine \emph{673}, \emph{676} and  \emph{679} have heavier online services (high memory usage), for the number of online container instances are 17, 19 and 18, respectively;
the machine \emph{1039} has a skew on the batch tasks  and  online container number, for the average number of batch tasks is 71, while the number of online container is 1.

\textbf{Summary}. In addition to system failures, unreasonable scheduling and workload imbalance are the main causes of anomalies in Alibaba’s cluster.

\section{Related work}

\textbf{Cluster trace studies}.
 In 2011, Google open-sourced the publicly available cluster trace data~\cite{google_traces}, which is a 29-day trace of over 25 million tasks across 12,500 heterogeneous machines.  And there are several works on analyzing Google trace from different perspectives.
 Reiss et al.~\cite{Google_trace_analysis} studied the heterogeneity and dynamicity properties of the Google workloads.
 Zhang et al. focused on characterizing run-time task resource usages of CPU, memory and disk~\cite{Google_usage_shapes}.
 Liu et al. focused on the frequency and pattern of machine maintenance events, job and task-level workload behaviors, and how the overall cluster resources are utilized~\cite{Google_workloads_analysis}.
 Di et al. focused on loads of jobs and machines, and compared the differences between a Google datacenter and other Grid/HPC systems~\cite{Google_grid}.
 Different from the Google trace, the Alibaba trace that was released in 2017, which contains information about the co-located container and batch workloads.
 Lu et al.~\cite{Alibaba_cluster_trace} performed characterization of the Alibaba trace to reveal the imbalance phenomena in cloud,
 such as spatial imbalance, temporal imbalance, imbalanced resource demands and utilization.
 Cheng et al.~\cite{Alibaba_Colocated_trace} focused on providing a unique and microscopic view about how the co-located
workloads interact and impact each other.
Liu et al.~\cite{Alibaba-Elasticity}  revealed that the resource allocation of the Alibaba semi-containerized co-location cluster achieves high elasticity and plasticity.
 In addition, some works also focus on the reliability analysis based on cluster traces, such as mining failure patterns~\cite{LogMaster}, failure prediction~\cite{Online_failure}, etc.
 Our study focuses on a unique view about node performance differences and anomalies in co-located workloads cluster.


\textbf{Cluster anomaly analysis}. A number of node comparison methods have been adopted for anomaly detection in large-scale clusters~\cite{Non-Parametric-Anomaly}. For example, most works use cosine similarity to calculate the node similarity in a cluster~\cite{BDTune}.
Kahuna \cite{Kahuna} aimed to diagnose performance based on node similarity, with supposing that nodes exhibit peer-similarity under fault-free conditions, and that some faults result in peer-dissimilarity.
Kasick et al.~\cite{Black-box-file} developed anomaly detection mechanisms in distributed environments by comparing system metrics among nodes.
Eagle~\cite{Eagle} is a framework for anomaly detection at eBay, which uses density estimation and PCA algorithms for user behavior analysis.

\section{Conclusion}

Aiming at improving the overall resource utilization, co-locating online services and offline batch jobs is an efficient approach.
However, it also results in exponentially increased complexity for datacenter resource management.
Based on the preprocessed  Alibaba co-located workloads dataset, we conducted in-depth analysis from the aspects of node similarity, workload characteristics and distribution, and anomalies.
 Our analysis reveals several insights that the performance discrepancy of machines in Alibaba's production cluster is relatively large, for
the distribution and resource utilization of co-located workloads are not balanced. 
For example,  the resource utilization (especially memory utilization) of  batch jobs  is fluctuating and  not as stable as that of online containers, and the reason is that the online containers are long-running jobs with more memory-demanding and most  batch jobs are short jobs.
 Meanwhile, based on the distribution of co-located workload instance numbers, the machines can be classified into 8 workload distribution
categories. And most patterns of machine resource utilization curves are similar in the same workload distribution category.
We also use the iForest algorithm to detect abnormal nodes, and find that the unreasonable scheduling and workload imbalance are the main causes of anomalies in Alibabas production cluster.


\section*{Acknowledgment}

We are very grateful to anonymous reviewers.

\bibliographystyle{ACM-Reference-Format}
\bibliography{references}

\end{document}